\documentclass[12pt]{article}

\usepackage[dvips]{color}                 
\usepackage{graphicx}                     
\usepackage{amssymb}
\usepackage{amsmath}

\usepackage{xspace}                       

\newcommand{\absatz}{\vspace{2ex}\noindent}

\renewcommand{\today}{\ifcase\day\or 1st\or 2nd\or 3rd\or 4th\or 5th\or 6th\or
  7th\or 8th\or 9th\or 10th\or 11th\or 12th\or 13th\or 14th\or 15th\or 16th\or
  17th\or 18th\or 19th\or 20th\or 21st\or 22nd\or 23rd\or 24th\or 25th\or
  26th\or 27th\or 28th\or 29th\or 30th\or 31st\fi~\ifcase\month\or January\or
  February\or March\or April\or May\or June\or July\or August\or September\or
  October\or November\or December\fi \space \number\year}

\newcommand{\journal}[4]{{#1}\textbf{#2},
  #4 (#3)}

\newcommand{\EPJA}{\textit{Eur.\  Phys.\ J.\ }\textbf{A}}

\newcommand{\NPA}{\textit{Nucl.\ Phys.\ }\textbf{A}}
\newcommand{\NPB}{\textit{Nucl.\ Phys.\ }\textbf{B}}
\newcommand{\PLB}{\textit{Phys.\ Lett.\ }\textbf{B}}
\newcommand{\PR}{\textit{Phys.\ Rev.\ }}

\newcommand{\PRC}{\PR\textbf{C}}
\newcommand{\PRD}{\PR\textbf{D}}
\newcommand{\PRL}{\PR\textit{Lett.\ }}

\newcommand{\HBChiPT}{HB$\chi$PT\xspace}

\newcommand{\D}{$\Delta(1232)$\xspace}
\newcommand{\Oeps}{$\mathcal{O}(\epsilon^3)$\xspace}

\newcommand{\calO}{\mathcal{O}}

\newcommand{\mytitle}[1]{\begin{center}\LARGE{\textbf{#1}}\end{center}}
\newcommand{\myauthor}[1]{\textbf{#1}}
\newcommand{\myaddress}[1]{\textit{#1}}
\newcommand{\mypreprint}[1]{\begin{flushright}#1\end{flushright}}

\begin{document}
%

\begin{titlepage}
  
  \mypreprint{ 
    \hfill
    nucl-th/0308054\\
    TUM-T39-03-15 }

  \vspace*{0.1cm}
  
  \mytitle{Spin Polarizabilities of the Nucleon from Polarized Low Energy
    Compton Scattering}

  \vspace*{0.3cm}

\begin{center}
  \myauthor{Robert P. Hildebrandt$^{a,}$}\footnote{Email: rhildebr@ph.tum.de},
  \myauthor{Harald W.\ Grie\3hammer$^{a,b,}$}\footnote{Email: hgrie@ph.tum.de;
    permanent address: a} \\
  and \myauthor{Thomas R.~Hemmert$^{a,b,}$}\footnote{Email:
    themmert@ph.tum.de; permanent address: a}\\[2ex]
  
  \vspace*{0.5cm}
  
  \myaddress{$^a$
    Institut f{\"u}r Theoretische Physik (T39), Physik-Department,\\
    Technische Universit{\"a}t M{\"u}nchen, D-85747 Garching, Germany}
  \\[2ex]
  \myaddress{$^b$ ECT*, Villa Tambosi, I-38050 Villazzano (Trento), Italy}


\end{center}

\vspace*{0.5cm}

\begin{abstract}
  As guideline for forthcoming experiments, we present predictions from Chiral
  Effective Field Theory
  for polarized cross sections in low energy Compton scattering for photon
  energies below $170$ MeV, both on the proton and on the neutron.
  Special interest is put on the role of the nucleon spin polarizabilities
  which can be examined especially well in polarized Compton scattering. We
  present a model-independent way to extract their energy dependence and
  static values from experiment, interpreting our findings also in terms of
  the low energy effective degrees of freedom inside the nucleon: 
The polarizabilities are dominated by chiral dynamics from the pion cloud, 
 except for resonant multipoles, where 
contributions of the $\Delta(1232)$
  resonance turn out to be crucial. We therefore include it as an explicit
  degree of freedom. We also identify some experimental settings which are
  particularly sensitive to the spin polarizabilities.

\end{abstract}
\vskip 1.0cm
\noindent
\begin{tabular}{rl}
Suggested PACS numbers:& 13.40.-f, 13.60.Fz, 14.20.Dh\\[1ex]
Suggested Keywords: &\begin{minipage}[t]{11cm}
                    Effective Field Theory, Polarized Compton Scattering,\\
                     Nucleon Spin Structure, Nucleon Polarizabilities. 
                    \end{minipage}
\end{tabular}

\vskip 1.0cm

\end{titlepage}

\setcounter{page}{2} \setcounter{footnote}{0} \newpage

%

\section{Introduction}
\setcounter{equation}{0}
\label{sec:introduction}

Over the past few decades, real Compton scattering off the proton was 
established as an
excellent tool to study the polarizabilities of the nucleon -- theoretically
as well as experimentally.  A good overview over the various experiments is
given in \cite{Pasquini}; for an overlook of the theoretical studies 
cf.  \cite{CD97, chiral2000} and
references therein. As is well-known, polarizabilities are a measure for the
stiffness of the nucleon in an external electric or magnetic field, caused by
the displacement of the charged constituents of the nucleon, induced by the
photon field.  While the static values $\bar{\alpha}_E$, $\bar{\beta}_M$ of the
two lowest (dipole) spin-independent polarizabilities are well understood,
there are only few experiments which are able to extract the \emph{spin
  polarizabilities} of the nucleon. These quantities have no simple classical
analogon, as they parameterize the stiffness of the nucleon spin against
electro-magnetically induced deformations relative to the axis defined by the
nucleon spin.
While there are four dipole spin polarizabilities
for each nucleon \cite{Ragusa}, the only two quantities measured
so far are the static forward and backward spin polarizabilities $\gamma_0$
and $\gamma_{\pi}$ of the proton.  $\gamma_0$ was extracted from the GDH
experiment at MAMI, using a Dispersion Relation (DR) analysis
\cite{Pasquini,GDH}:
\begin{equation}
  \gamma_0=(-1.01\pm0.13)\cdot10^{-4}\;\mathrm{fm}^4
\label{eq:gamma0}
\end{equation}
A first attempt to determine $\gamma_{\pi}$ from experiment by the
LEGS group \cite{Sandorfi} quotes
\begin{equation}
  \gamma_\pi=(-27.1\pm3.6)\cdot10^{-4}\;\mathrm{fm}^4,
\label{LEGS}
\end{equation}
which is considerably lower than what one expected from DR analysis and Chiral
Effective Field Theory ($\chi$EFT). An extraction from recent MAMI data,
obtained at low energies \cite{Olmos01} and in the region of the $\Delta$
resonance \cite{Galler, Wolf, Camen}, yields values which differ strongly (on
a level of about 30\%) from the LEGS value: 
\begin{align}
  \gamma_\pi&=(-36.1\pm2.2)\cdot10^{-4}\;\mathrm{fm}^4\;\;\;\cite{Olmos01}
  \nonumber\\
  \gamma_\pi&=(-37.9\pm3.6)\cdot10^{-4}\;\mathrm{fm}^4\;\;\;\cite{Galler}
\end{align}
These new results agree very well with the theoretical prediction from
$\chi$EFT, 
\begin{align}
  \gamma_\pi&=-36.7\cdot10^{-4}\;\mathrm{fm}^4\;\;\;\cite{HHKK},
\label{eq:gammapi}
\end{align}
whereas calculations based on $\chi$EFT are at present not able to reproduce
the MAMI value for $\gamma_0$, Eq.~(\ref{eq:gamma0})\footnote{In the forward
  direction, a strong cancellation between two large spin polarizabilities
  makes accurate predictions for $\gamma_0$ rather difficult
  \cite{chiral2000}.}.  For further details concerning experiments and their
results see e.g.~\cite{Pasquini}.

The goal of this work is to motivate further investigations of the spin
polarizabilities, where there are still so many question marks left.
Especially, we advocate double polarized experiments as a tool to dis-entangle
the four leading spin polarizabilities, and not only the two static linear
combinations given above.

Recently, it was demonstrated in \cite{GH01} that nucleon polarizabilities can
 be connected to Compton multipoles and therefore also acquire a dependence 
on the energy
$\omega$ of the real, incoming photon. These dispersion effects are  
well-known in solid state physics.
In hadron stucture physics different internal nucleonic degrees of freedom,
low-lying nuclear resonances like the $\Delta(1232)$, the charged meson cloud
around the nucleon etc., will react quite differently to real photon fields of
non-zero frequency.  Therefore, these \emph{dynamical polarizabilities}
contain detailed information about dispersive effects, caused by internal
relaxation effects, baryonic resonances and mesonic production thresholds, see
\cite{GH01, HGHP03} for details.  As they stem from a multipole analysis of
the scattering amplitude, dynamical polarizabilities contain all hadron 
structure information, 
but in a more readily accessible form. In the limit of zero photon
energy, they reduce to the static polarizabilities mentioned above.

In principle, the dynamical polarizabilities are experimentally accessible by
fits to Compton scattering cross sections. The main problem seems to be that
the multipole expansion allows for an a priori infinite number of fit
functions: The real photons can undergo transitions $Tl\to T^\prime l^\prime$,
where $T/T^\prime=E$ or $M$ labels the coupling of the incoming/outgoing
photon as electric or magnetic, and $l$, ($l^\prime=l\pm\{0;1\}$) is the
angular momentum of the incident (outgoing) photon.  Thus, there are six,
$\omega$-dependent dipole polarizabilities, namely the two spin-independent
ones $\alpha_{E1}(\omega)$ and $\beta_{M1}(\omega)$ for electric and magnetic
dipole transitions which do not couple to the nucleon spin. In the spin sector
 there are the two diagonal polarizabilities $\gamma_{E1E1}(\omega)$ and 
$\gamma_{M1M1}(\omega)$ 
and the two off-diagonal spin polarizabilities $\gamma_{E1M2}(\omega)$ and
$\gamma_{M1E2}(\omega)$. 
In addition,
there are higher ones like quadrupole and octupole polarizabilities.
In \cite{HGHP03}, however, it was shown that one can describe unpolarized 
low energy
Compton scattering off the proton very well by keeping only the $l=1$ (dipole)
contributions of the Compton multipoles. This leaves us with six unknown
functions of the photon energy that can be expressed as the six dynamical
dipole polarizabilities. 

Obviously, further experiments are needed to 
determine these six functions,
as there is e.g.~only a minor dependence on the spin polarizabilities visible 
in spin-averaged cross sections below the pion production threshold (see Sect.
\ref{spinav}).  Polarized Compton scattering experiments provide a new avenue
for the determination of the six dipole polarizabilities.  In the seminal
paper \cite{Babusci} on polarized Compton scattering off a nucleon, an
exhaustive list of interesting observables and asymmetries was defined which
only now start to become accessible in this new frontier of low energy
electromagnetic scattering experiments.

Guided by ongoing experimental feasibility studies at the HI$\gamma$S lab of
TUNL \cite{Gao}, we chose a subset of four asymmetries describing the
interaction of circularly polarized photons with polarized protons and
neutrons, where the polarization in the final states is not detected. We cover
the low energy range, up to photon energies of $\sim 170$ MeV, just above the
one pion production threshold. Like in \cite{Babusci}, we focus on
asymmetries, dividing the difference of two polarized cross sections by their
sum, as they are less sensitive to experimental errors than differences.
Further investigations involving linearly polarized photon beams are under
study \cite{linear}.
We present predictions in the framework of Chiral Effective Field Theory with
explicit $\Delta$ degrees of freedom.  Previously, a calculation of two of the
asymmetries for polarized Compton scattering of the proton was presented to
leading-one-loop order in a $\chi$EFT with only nucleon and pion degrees of
freedom in \cite{BKM}. In our analysis, (Sects.~\ref{sec:proton}
and~\ref{sec:neutron}) we show a comparison between the two chiral frameworks
for all asymmetries we consider, so that $\Delta$ physics is easily
identifiable.
Since we investigate the possibility of determining spin polarizabilities from
experiment, we put special emphasis on the role of the spin and quadrupole
($l=2$) polarizabilities of the nucleon. The latter ones will -- as in
\cite{HGHP03} for the spin-averaged case -- turn out to be negligibly small,
leaving only the six dynamical dipole polarizabilities as unknown structure
parameters to be determined from data.
As a starting point, one might consequently accept the theoretical findings
for the spin-independent dipole polarizabilities $\alpha_{E1}(\omega)$,
$\beta_{M1}(\omega)$, for which $\chi$EFT and Dispersion analysis are in good
agreement \cite{HGHP03}. One would then attempt to extract the four dynamical 
$l=1$ spin polarizabilities directly from experiment, as will be described in
Sect.~\ref{strucamp}.

In \cite{Babusci}, the authors also investigated the energy dependence of
various asymmetries both in a low energy expansion in terms of nucleon
polarizabilities as well as in a full calculation in Dispersion Theory.  The
low energy expansion included the static values of the six dipole
polarizabilities,
the two static spin-independent quadrupole polarizabilities
$\bar{\alpha}_{E2},\;\bar{\beta}_{M2}$, and the leading dispersion corrections
to $\bar{\alpha}_{E1}$ and $\bar{\beta}_{M1}$. Such a Taylor expansion of the
polarizabilities is bound to break down as cusps or resonances are approached,
the lowest of which being the one pion production threshold and the $\Delta$
resonance. We will indeed find strong signals from the spin polarizabilities as
the energy is increased. In contradistinction to \cite{Babusci}, our
calculation is based on \textit{dynamical}, i.e.~energy-dependent
polarizabilities \cite{GH01}.  
For photon energies above the pion production threshold, we expect our
predictions only qualitatively  to be correct,
since the imaginary parts of our dynamical polarizabilities only correspond to
tree-level accuracy and the width of the $\Delta(1232)$ resonance is treated
as a small perturbation; for details see \cite{HGHP03}.  Whereas the
unpolarized cross sections were found to be well described up to $170$ MeV
within our theoretical framework \cite{HGHP03}, we have to caution the reader 
that this does not have to be the case for the asymmetries presented here,
as these are much more sensitive on fine details.

The two spin configurations we investigate are described in
Sect.~\ref{sec:form}, after a short repetition of the theoretical framework
(Sect.~\ref{sec:theoretical_framework}). In Sect.~\ref{strucamp}, we propose a
procedure to determine spin polarizabilities from experiment, before we have
another look at spin-averaged cross sections in Sect.~\ref{spinav}.  The
results for the proton asymmetries are presented and interpreted in
Sect.~\ref{sec:proton}, the ones for the neutron in Sect.~\ref{sec:neutron}.
Conclusions and an Appendix on the two dominating Compton amplitudes complete
the presentation. 

\section{Theoretical Framework}
\setcounter{equation}{0}
\label{sec:theoretical_framework}

We calculate in the framework of Chiral Effective Field Theory with and
without explicit $\Delta(1232)$ degrees of freedom. Details concerning the
former, Heavy Baryon Chiral Perturbation Theory (\HBChiPT), which contains
only pions and nucleons as explicit degrees of freedom, can be found e.g.~in
\cite{BKM}. The formalism of the latter, called ``Small Scale Expansion''
(SSE) -- an effective chiral field theory describing explicit pion, nucleon
and \D degrees of freedom -- is discussed in \cite{HHK98}.  This work is based
on the calculation of dynamical nucleon polarizabilities and spin-averaged
Compton cross sections of the proton in \cite{HGHP03} to which we refer the
interested reader for details of our notation.


Real Compton scattering can be formulated in terms of six
amplitudes\footnote{These amplitudes are different from the amplitudes $A_i$
  in \cite{Babusci}, as we use a different basis.} $A_1-A_6$. The $T$-matrix
reads
\begin{equation}
\begin{split}
  T(\omega,z)&= A_1(\omega,z)\,\vec{\epsilon}\,'^*\cdot \vec{\epsilon} +
  A_2(\omega,z)\,
  \vec{\epsilon}\,'^*\cdot\hat{\vec{k}}\,\vec{\epsilon}\cdot\hat{\vec{k}}'\\
  &+i\,A_3(\omega,z)\,\vec{\sigma}\cdot
  \left(\vec{\epsilon}\,'^*\times\vec{\epsilon}\,\right)
  +i\,A_4(\omega,z)\,\vec{\sigma}\cdot
  \left(\hat{\vec{k}}'\times\hat{\vec{k}}\right)
  \vec{\epsilon}\,'^*\cdot\vec{\epsilon}\\
  &+i\,A_5(\omega,z)\,\vec{\sigma}\cdot
  \left[\left(\vec{\epsilon}\,'^*\times\hat{\vec{k}} \right)\,
    \vec{\epsilon}\cdot\hat{\vec{k}}' -\left(\vec{\epsilon}
      \times\hat{\vec{k}}'\right)\,
    \vec{\epsilon}\,'^*\cdot\hat{\vec{k}}\right]\\
  &+i\,A_6(\omega,z)\,\vec{\sigma}\cdot
  \left[\left(\vec{\epsilon}\,'^*\times\hat{\vec{k}}'\right)\,
    \vec{\epsilon}\cdot\hat{\vec{k}}' -\left(\vec{\epsilon}
      \times\hat{\vec{k}} \right)\,
    \vec{\epsilon}\,'^*\cdot\hat{\vec{k}}\right]
\end{split} 
\label{Tmatrix}
\end{equation}
with $\hat{\vec{k}}$ ($\hat{\vec{k}}'$) the unit vector in the momentum
direction of the incoming (outgoing) photon with polarization $\vec{\epsilon}$
($\vec{\epsilon}\,'^*$).  We separate these amplitudes into pole
($A_i^\mathrm{pole}$) and non-pole ($\bar{A}_i$)
parts.

The non-pole amplitudes are also referred to as the structure part of
the amplitudes. The question whether a contribution belongs to the structure
part cannot be answered uniquely. In our definition, only those terms which
have a pole either in the $s$-, $u$- or $t$-channel are treated as
non-structure. 
If we were only concerned with the full calculation of Compton scattering
cross sections, this separation clearly would be irrelevant because both, the
structure dependent as well as the structure independent part, contribute.
Here, however, we investigate the role of the internal nucleonic degrees of
freedom on the spin and quadrupole polarizabilities in Compton scattering.
Therefore, we need to be able to turn off and on the different nucleon
polarizabilities, which are contained only in the structure part of the
amplitudes.

Expressing the $l=1$ multipole expansion for nucleon Compton scattering in 
terms of dynamical dipole polarizabilities, one obtains
\begin{align}
  \bar{A}_1(\omega,\,z) &=\frac{4\pi\,W}{M}\,\left[\alpha_{E1}(\omega)
    +z\,\beta_ {M1}(\omega)\right]\,\omega^2+\mathcal{O}(l=2),\nonumber\\
  \bar{A}_2(\omega,\,z) &=-\frac{4\pi\,W}{M}\,\beta_{M1}(\omega)\,\omega^2
  +\mathcal{O}(l=2),\nonumber\\
  \bar{A}_3(\omega,\,z) &=-\frac{4\pi\,W}{M}\,\left[\gamma_{E1E1}(\omega)
    +z\,\gamma_{M1M1}(\omega)+\gamma_{E1M2}(\omega)
    +z\,\gamma_{M1E2}(\omega)\right]\,\omega^3+\mathcal{O}(l=2),\nonumber\\
  \bar{A}_4(\omega,\,z) &=\frac{4\pi\, W}{M}\,\left[-\gamma_{M1M1}(\omega)
    +\gamma_{M1E2}(\omega)\right]\,\omega^3+\mathcal{O}(l=2),\nonumber\\
  \bar{A}_5(\omega,\,z) &=\frac{4\pi\, W}{M}\,\gamma_{M1M1}(\omega)\,\omega^3
  +\mathcal{O}(l=2),\nonumber\\
  \bar{A}_6(\omega,\,z) &=\frac{4\pi\,
    W}{M}\,\gamma_{E1M2}(\omega)\,\omega^3+\mathcal{O}(l=2).
\label{eq:strucamp}
\end{align}
We choose to work in the centre-of-mass frame. Thus, $\omega$ denotes the cm
energy of the photon, M the isoscalar nucleon mass, $W=\sqrt{s}$ the total
cm-energy, and $\theta$ the cm-scattering angle with $z=\cos\theta$.

The structure amplitudes $\bar{A}_3-\bar{A}_6$ contain only spin
polarizabilities, $\bar{A}_1-\bar{A}_2$ only spin-independent ones.  The spin
polarizabilities $\gamma_0\;(\gamma_\pi)$ mentioned in the introduction are
the leading coefficients of $\bar{A}_3$ for
$\theta=0^\circ\,\;(\theta=180^\circ)$ at zero energy:
$$
\gamma_0 =-(\bar{\gamma}_{E1E1}+\bar{\gamma}_{M1M1}+
\bar{\gamma}_{E1M2}+\bar{\gamma}_{M1E2}),\;\; \gamma_\pi=-
\bar{\gamma}_{E1E1}+\bar{\gamma}_{M1M1}-
\bar{\gamma}_{E1M2}+\bar{\gamma}_{M1E2}
$$
and do thus not suffice to determine the four leading spin polarizabilities
completely.  Here, $\bar{\gamma}_i$ denotes the static limit
$\bar{\gamma}_i=\gamma_i(\omega=0)$.  A precise definition of polarizabilities
via the multipole expansion of the amplitudes is given in \cite{HGHP03}.


\absatz
In the following, we list all the diagrams contributing in our 
leading-one-loop order
(\Oeps) calculation in the Small Scale Expansion, which contains the leading
chiral dynamics of the pion cloud and the dominant $\Delta$ physics with its
pionic cloud. $\epsilon$ is the expansion parameter of SSE and denotes either
a small momentum, the pion mass or the mass difference between nucleon and
$\Delta(1232)$.  A diagram at a certain order in $p$ containing pions in the
theory without explicit $\Delta$ degrees of freedom, \HBChiPT, contributes at
the same order $\epsilon$ in SSE.

\begin{figure}[!htb]
\begin{center} 
  \includegraphics*[width=.8\textwidth] {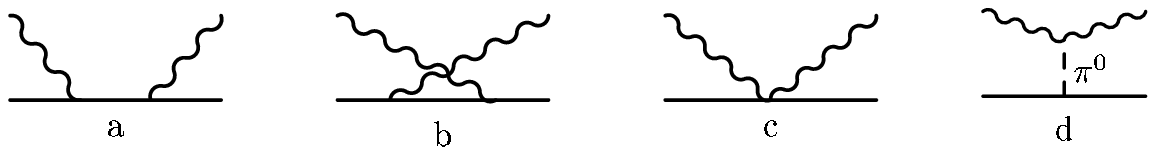}
\caption{The diagrams identified as pole contributions at
  leading-one-loop order in \HBChiPT.
}
\label{poldiagramme}
\end{center}
\end{figure}

In Fig.~\ref{poldiagramme}, we show the four \HBChiPT non-structure (pole)
diagrams which contribute to an $\mathcal{O}(p^3)$- (and therefore also to an
$\mathcal{O}(\epsilon^3)$-) calculation: the pole diagrams (a,b), the Thomson
term (c) and the ``pion pole'' (d). The pole parts are thus given by the
amplitudes of Compton scattering off a point-like nucleon with an anomalous
magnetic moment, in addition to the $\pi^0$-pole contribution in the 
$t$-channel. In
the literature, the latter contribution is sometimes classified as a structure
part. 

To $\mathcal{O}(\epsilon^3)$,  one obtains now for the proton non-structure
amplitudes
\begin{align}
  A_1^{\mathrm{pole,\,p}}(\omega,\,\theta)&=-\frac{e^2}{M}
  +\mathcal{O}(\epsilon^4),  \nonumber\\
  A_2^{\mathrm{pole,\,p}}(\omega,\,\theta)&= \frac{e^2\,\omega}{M^2}
  +\mathcal{O}(\epsilon^4),  \nonumber\\
  A_3^{\mathrm{pole,\,p}}(\omega,\,\theta)&=
  \frac{e^2\,\omega\,\left(1+2\,\kappa_p-(1+\kappa_p)^2\,\cos\theta\right)}
  {2\,M^2} -\frac{e^2\,g_A}{4\,\pi^2\,f_\pi^2}\,
  \frac{\omega^3\,(1-\cos\theta)}{m_{\pi^0}^2+2\, \omega^2\,(1-\cos\theta)}
  +\mathcal{O}(\epsilon^4),\nonumber\\
  A_4^{\mathrm{pole,\,p}}(\omega,\,\theta)&=
  -\frac{e^2\,\omega\,(1+\kappa_p)^2} {2\,M^2}
  +\mathcal{O}(\epsilon^4),\nonumber\\
  A_5^{\mathrm{pole,\,p}}(\omega,\,\theta)&=
  \frac{e^2\,\omega\,(1+\kappa_p)^2} {2\,M^2}
  -\frac{e^2\,g_A}{8\,\pi^2\,f_\pi^2}\,\frac{\omega^3}{m_{\pi^0}^2+2\,
    \omega^2\,(1-\cos\theta)}
  +\mathcal{O}(\epsilon^4),  \nonumber\\
  A_6^{\mathrm{pole,\,p}}(\omega,\,\theta)&=
  -\frac{e^2\,\omega\,(1+\kappa_p)}{2\,M^2}
  +\frac{e^2\,g_A}{8\,\pi^2\,f_\pi^2}\,\frac{\omega^3}{m_{\pi^0}^2+2\,
    \omega^2\,(1-\cos\theta)} +\mathcal{O}(\epsilon^4).
\label{eq:pole}
\end{align}


The neutron pole amplitudes are
\begin{align}
  A_1^{\mathrm{pole,\,n}}(\omega,\,\theta)&=0+\mathcal{O}(\epsilon^4),
  \nonumber\\
  A_2^{\mathrm{pole,\,n}}(\omega,\,\theta)&=0+\mathcal{O}(\epsilon^4),
  \nonumber\\
  A_3^{\mathrm{pole,\,n}}(\omega,\,\theta)&=
  -\frac{e^2\,\omega\,\kappa_n^2\,\cos\theta}{2\,M^2}
  +\frac{e^2\,g_A}{4\,\pi^2\,f_\pi^2}\,
  \frac{\omega^3\,(1-\cos\theta)}{m_{\pi^0}^2+2\,\omega^2\,(1-\cos\theta)}
  +\mathcal{O}(\epsilon^4), \phantom{111111111111111}
  \nonumber\\
  A_4^{\mathrm{pole,\,n}}(\omega,\,\theta)&=-\frac{e^2\,\omega\,\kappa_n^2}
  {2\,M^2}+\mathcal{O}(\epsilon^4),\nonumber\\
  A_5^{\mathrm{pole,\,n}}(\omega,\,\theta)&= \frac{e^2\,\omega\,\kappa_n^2}
  {2\,M^2} +\frac{e^2\,g_A}{8\,\pi^2\,f_\pi^2}\,\frac{\omega^3}
  {m_{\pi^0}^2+2\,\omega^2\,(1-\cos\theta)} +\mathcal{O}(\epsilon^4),
  \nonumber\\
  A_6^{\mathrm{pole,\,n}}(\omega,\,\theta)&=
  -\frac{e^2\,g_A}{8\,\pi^2\,f_\pi^2}\,\frac{\omega^3}
  {m_{\pi^0}^2+2\,\omega^2\,(1-\cos\theta)} +\mathcal{O}(\epsilon^4).
\label{eq:Bornn}
\end{align}
$\kappa_p$ ($\kappa_n$) is the anomalous magnetic moment of the proton
(neutron), $e$ the proton's electric charge.  $m_{\pi^0}$ is the mass of the
neutral pion, $f_\pi$ the pion decay constant.  The terms containing the axial
coupling constant $g_A$ are the contributions of the pion pole. The numerical
values we use are listed in Table~\ref{table}.

\begin{table}[!htb]
\begin{center}
\begin{tabular}{|c|c|c||c|}
\hline 
Parameter & Value & Comment&Source \\
\hline 
$m_{\pi^0}$& $135.0$ MeV &neutral pion mass&\cite{PDG}\\
$m_\pi$  & $139.6$ MeV &charged pion mass&\cite{PDG} \\
$M$ & $938.9$ MeV &isoscalar nucleon mass&\cite{PDG} \\
$f_\pi$  & $92.4$ MeV &pion decay constant&\cite{PDG}\\
$g_A$    & $1.267$&axial coupling constant&\cite{PDG} \\
$e$      & $\sqrt{4\,\pi/137}$&electric charge of the proton&\cite{PDG}\\
$\kappa_p$ & $1.793$ & anom.~mag.~moment proton&\cite{PDG} \\
$\kappa_n$ & $-1.913$ & anom.~mag.~moment neutron&\cite{PDG} \\
\hline
\end{tabular}
\caption{Numerical values; magnetic moments are given in nuclear magnetons.}
\label{table}
\end{center}
\end{table}

\begin{figure}[!htb]
  \begin{center}
    \includegraphics*[width=.8\textwidth] {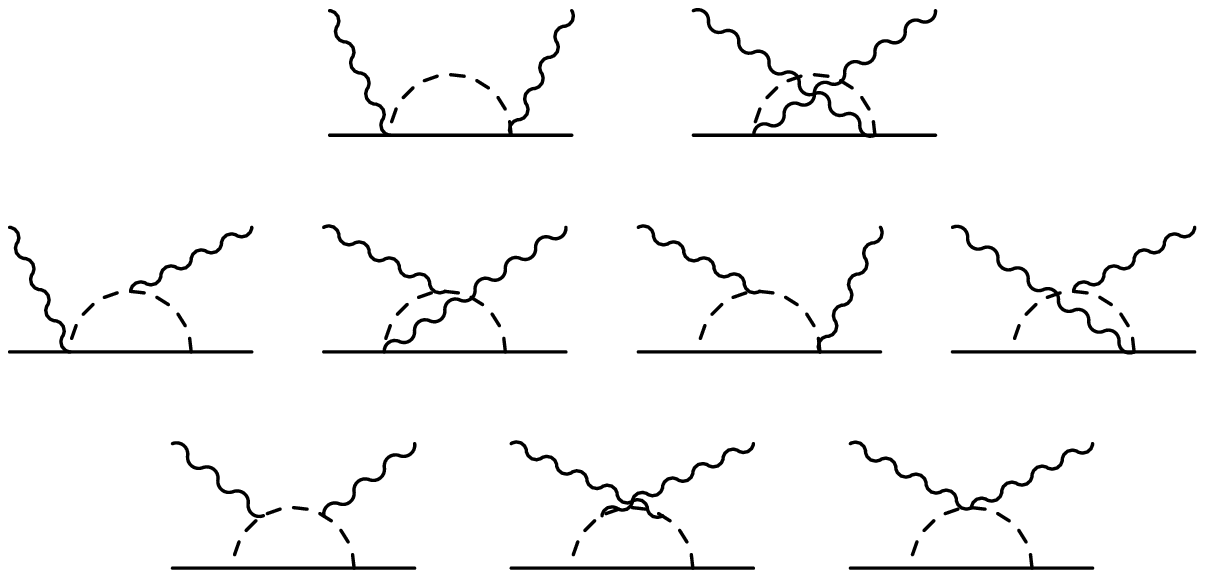}
    \caption{All diagrams contributing to the structure dependent amplitudes at
      leading-one-loop order in \HBChiPT.}
    \label{NpiKontinuum}
  \end{center}
\end{figure}

\begin{figure}[!htb]
  \begin{center}
    \includegraphics*[width=.8\textwidth] {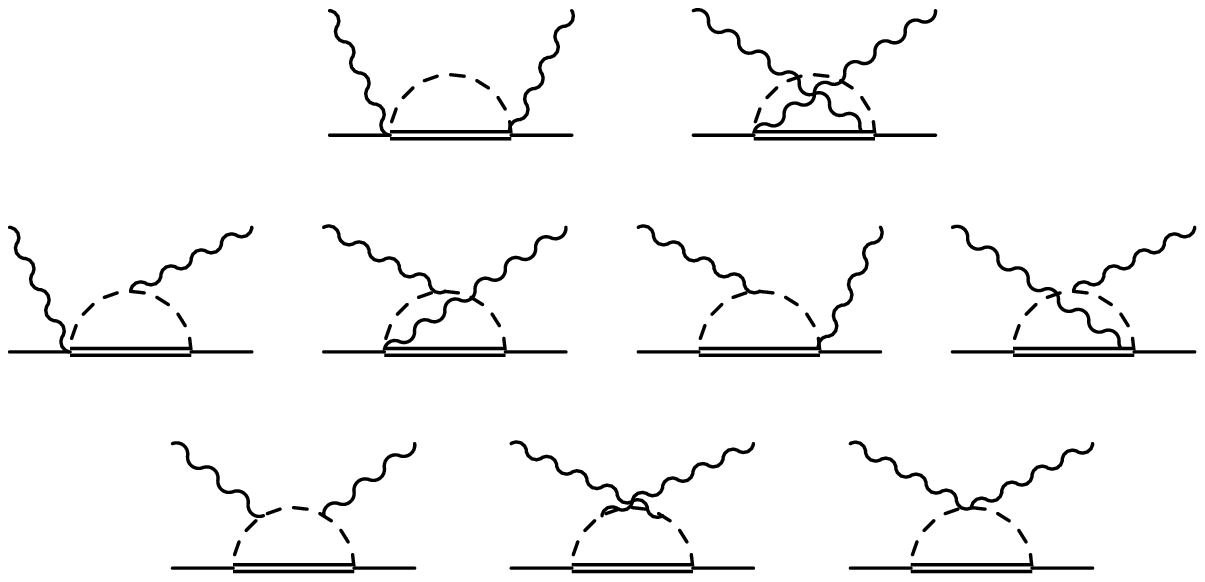}
    \caption{The $\Delta$-$\pi$-continuum diagrams contributing at 
  leading-one-loop order in SSE. The double line denotes the $\Delta(1232)$. }
 \label{DeltapiKontinuum}  
\end{center}
\end{figure}

\begin{figure}[!htb]
\begin{center}
  \includegraphics*[width=.65\textwidth] {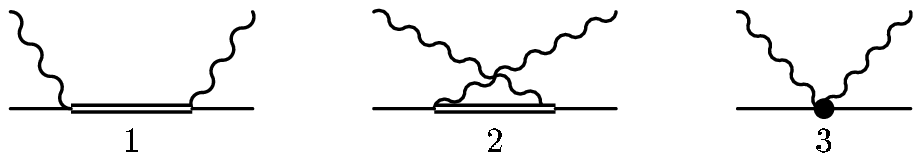}
\caption{$\Delta$ pole diagrams at leading-one-loop order and seagull graph.} 
\label{Deltapoldiagramme}
\end{center}
\end{figure}

Fig.~\ref{NpiKontinuum} shows the non-pole terms in leading-one-loop order
HB$\chi$PT~\cite{BKKM}. To this order, they contain only pion cloud effects
around the nucleon.  In SSE, the diagrams at order $\epsilon^3$ in addition to
the \HBChiPT ones of Figs.~\ref{poldiagramme} and \ref{NpiKontinuum} are the
$\Delta$-$\pi$-continuum~(Fig.~\ref{DeltapiKontinuum}), and $\Delta$ pole
diagrams
(Figs.~\ref{Deltapoldiagramme}.1,~\ref{Deltapoldiagramme}.2)~\cite{HHKK,HHK97}.
We emphasize that there is no difference in the structure part of the
amplitudes between proton and neutron up to $\mathcal{O}(\epsilon^3)$.
Therefore, our non-pole amplitudes describe an isoscalar nucleon and the only
difference between the two nucleons comes in via the pole amplitudes
Eqs.~(\ref{eq:pole}, \ref{eq:Bornn}).  As the analytic expressions for the
structure amplitudes are rather lengthy, we refer the reader to \cite{HGHP03}
for a complete listing.


As discussed in \cite{HGHP03} we require two additional operators --
represented by the seagull graph in Fig.~\ref{Deltapoldiagramme}.3 -- which
are energy-independent and contribute only to the spin-independent dipole
polarizabilities.  These terms are formally $\mathcal{O}(\epsilon^4)$, but
turn out to be anomalously large \cite{HGHP03} and therefore have to be taken
into account already at leading-one-loop order. Their numerical values are
determined by a fit to unpolarized Compton scattering data. The static, 
spin-independent dipole polarizabilities $\bar{\alpha}_E$ and $\bar{\beta}_M$ 
thus obtained are in excellent agreement to and of comparable uncertainty as 
the results from alternative extractions, see \cite{HGHP03} for details. 

We now turn to the formalism of Compton cross sections.


\section{Cross Sections and Asymmetries -- Formalism}
\setcounter{equation}{0}
\label{sec:form}

The well-known formula for Compton cross sections in the cm frame is
\begin{equation}
  \left.\frac{d\sigma}{d\Omega}\right|_\mathrm{cm}=
  \left(\frac{M}{4\pi\,W}\right)^2\; |T|^2\;\;.
\label{eq:diffcrosssection}
\end{equation}

%
%


In \cite{HGHP03}, we showed results for unpolarized proton cross sections,
which are derived by averaging over the initial and summing over the final
spin states. Now, we concentrate on spin-polarized cross sections for proton
and neutron, albeit we will return briefly to spin-averaged ones in
Sect.~\ref{spinav}.

Triggered by a forthcoming proposal on polarized Compton scattering off
$\,^3\mathrm{He}$ at the HI$\gamma$S lab of TUNL \cite{Gao}, we choose the
incoming photon to be right-circularly polarized,
$$\vec{\epsilon}= \frac{1}{\sqrt{2}}\,
\begin{pmatrix}
  1\\
  i\\
  0
\end{pmatrix}$$
and moving along the positive $z$-direction, while the final polarization and
nucleon spin remain undetected.  The two nucleon spin configurations we
investigate are:
\begin{itemize}
\item[1)] the difference between the target nucleon spin pointing parallel or
  antiparallel to the incident photon momentum
  $$\frac{d\sigma_{\uparrow\uparrow }}{d\Omega_{\mathrm{cm}}} -
  \frac{d\sigma_{\uparrow\downarrow}}{d\Omega_{\mathrm{cm}}};$$
  
\item[2)] the difference between the target nucleon spin aligned in positive
  or negative $x$-direction:
  $$\frac{d\sigma_{\uparrow\rightarrow}}{d\Omega_{\mathrm{cm}}} -
  \frac{d\sigma_{\uparrow\leftarrow }}{d\Omega_{\mathrm{cm}}}.$$
\end{itemize}

The corresponding formulae for $|T|^2$ can already be found in \cite{BKM},
albeit there they are given only for real amplitudes $A_1-A_6$. As is 
well-known, these amplitudes become complex for a photon energy above the pion
production threshold $\omega_\pi$.  Including the imaginary part of the
amplitudes, the formulae read
\begin{equation}
\begin{split}
  \frac{1}{2}\,(|T|^2_{\uparrow\uparrow}&-|T|^2_{\uparrow\downarrow})
  =-\mathrm{Re}[A_1\,A_3^*]\,(1+\cos^2\theta)
  -\biggl[|A_3|^2+2\,|A_6|^2+2\,|A_5|^2\,\cos^2\theta\\
  &+\mathrm{Re}[A_6\,(A_1^*+3\,A_3^*)]
  +\biggl(\mathrm{Re}[A_3\,(3\,A_5^*+A_4^*-A_2^*)]
  +\mathrm{Re}[A_5\,(4\,A_6^*-A_1^*)]\biggl)\,\cos\theta\\
  &+\mathrm{Re}[A_5\,(A_2^*-A_4^*)]\,\sin^2\theta \biggl]\,\sin^2\theta
\end{split}
\label{eq:paranti}
\end{equation}
and
\begin{equation}
\begin{split}
  \frac{1}{2}\,(|T|^2_{\uparrow\rightarrow}&-|T|^2_{\uparrow\leftarrow})
  =\biggl[ \mathrm{Im}[A_1\,(A_3^*+2\,A_6^*+2\,A_5^*\,\cos\theta)]\,\cos\theta
  +\mathrm{Im}[A_1\,A_4^*]\,(1+\cos^2\theta)\\
  &-\mathrm{Im}[A_2\,(A_3^*+2\,A_6^*)]\,\sin^2\theta
  -\mathrm{Im}[A_2\,(A_4^*+2\,A_5^*)]\,\cos\theta\,\sin^2\theta\biggl]
  \sin\theta\,\sin\phi\\
  &+\biggl[\mathrm{Re}[A_3\,(A_3^*-A_1^*+2\,A_6^*)]\,\cos\theta
  +\mathrm{Re}[A_3\,A_5^*]\,(3\,\cos^2\theta-1)\\
  &+\biggl(\mathrm{Re}[A_1\,A_5^*]+\mathrm{Re}[A_2\,A_3^*]
  +\mathrm{Re}[A_6\,(A_2^*+A_4^*-2\,A_5^*)]\biggl)\,\sin^2\theta\\
  &+\mathrm{Re}[A_3\,A_4^*]\,(\cos^2\theta+1)
  +\mathrm{Re}[A_5\,(A_2^*-A_4^*-2\,A_5^*)]\,\cos\theta\,\sin^2\theta
  \biggl]\sin\theta\,\cos\phi\;\;.
\end{split}
\label{eq:rightleft}
\end{equation}
Here, $\phi$ is the angle between the reaction plane and the plane spanned by
the momentum of the incoming photon and the target nucleon spin. Obviously,
the difference Eq.~(\ref{eq:rightleft}) takes on the largest values -- at
least below the pion production threshold -- for $\phi=0$. Therefore, we
choose the nucleon spin in the reaction plane, which simplifies
Eq.~(\ref{eq:rightleft}) considerably.  Using left- instead of
right-circularly polarized photons changes only the overall sign in
Eq.~(\ref{eq:paranti}) and Eq.~(\ref{eq:rightleft}).

For comparison, we show once again $|T|^2$ for the spin-averaged cross section
\cite{BKM}, which can be derived by taking the sum instead of the difference
in Eq.~(\ref{eq:paranti}) (or as well in Eq.~(\ref{eq:rightleft})):
\begin{equation}
\begin{split}
  \frac{1}{2}\,(|T|^2_{\uparrow\uparrow}&+|T|^2_{\uparrow\downarrow})
  =\frac{1}{2}\,|A_1|^2\,(1+\cos^2\theta)+\frac{1}{2}\,|A_3|^2\,
  (3-\cos^2\theta)
  +\biggl[\frac{1}{2}\,|A_4|^2\,(1+\cos^2\theta)\\
  &+\frac{1}{2}\,|A_2|^2\,\sin^2\theta+|A_5|^2\,(1+2\,\cos^2\theta)
  +3\,|A_6|^2+4\,\mathrm{Re}[A_3\,A_6^*]
  +2\,\mathrm{Re}[A_4\,A_5^*]\,\cos^2\theta\\
  &+\biggl(-\mathrm{Re}[A_1\,A_2^*] +\mathrm{Re}[A_3\,(A_4^*+2\,A_5^*)]
  +2\,\mathrm{Re}[A_6\,(A_4^*+3\,A_5^*)]\biggl)\,\cos\theta
  \biggl]\,\sin^2\theta
\end{split}
\label{eq:sum}
\end{equation}

The asymmetries we consider\footnote{$\Sigma_z$ corresponds to $\Sigma_{2z}$
  in the notation of \cite{Babusci}, $\Sigma_x$ to $\Sigma_{2x}$.} are
\begin{equation}
\Sigma_z=\frac{|T|^2_{\uparrow\uparrow}-|T|^2_{\uparrow\downarrow}}
              {|T|^2_{\uparrow\uparrow}+|T|^2_{\uparrow\downarrow}}\;\;,
\end{equation}
\begin{equation}
\Sigma_x=\frac{|T|^2_{\uparrow\rightarrow}-|T|^2_{\uparrow\leftarrow}}
              {|T|^2_{\uparrow\rightarrow}+|T|^2_{\uparrow\leftarrow}}\;\;.
\end{equation}
$\Sigma$ is a frame independent quantity, as the frame dependent flux factor
cancels in the ratio between difference and sum of the cross section, while
$|T|^2$ can be written in terms of the frame independent Mandelstam variables.

From the experimentalist's point of view, it is more convenient to measure the
\textit{asymmetry} -- i.e.~the difference divided by the sum -- instead of the
differences Eq.~(\ref{eq:paranti}) and Eq.~(\ref{eq:rightleft}), as the former
is more tolerant to systematic errors in experiments.  Nevertheless, division
by a small quantity, say a small spin-averaged cross section, may enhance
theoretical uncertainties.  Sensitivity on the nucleon structure, e.g.~the
nucleon spin, may be lost by dividing the difference by the sum, as we will
see a few times in Sects. \ref{sec:proton} and \ref{sec:neutron}.
We therefore give also two more definitions which abbreviate the cm
differences:
\begin{equation}
\mathcal{D}_z=\frac{1}{2}\,\left(
  \frac{d\sigma_{\uparrow\uparrow  }}{d\Omega_{\mathrm{cm}}}
 -\frac{d\sigma_{\uparrow\downarrow}}{d\Omega_{\mathrm{cm}}}\right)
\end{equation}
\begin{equation}
\mathcal{D}_x=\frac{1}{2}\,\left(
  \frac{d\sigma_{\uparrow\rightarrow}}{d\Omega_{\mathrm{cm}}}
 -\frac{d\sigma_{\uparrow\leftarrow }}{d\Omega_{\mathrm{cm}}}\right)
\end{equation}

\section{Extracting Spin Polarizabilities from Experiment}
\setcounter{equation}{0}
\label{strucamp}


A first step in determining dynamical spin polarizabilities from experiment
might be to accept our findings for the spin-independent dipole
polarizabilities $\alpha_{E1}(\omega)$ and $\beta_{M1}(\omega)$, which show
very good agreement with Dispersion Relation analysis up to about $170$ MeV
\cite{HGHP03}. 
Truncating at $l=1$, this
leaves no unknowns in $A_1$ and $A_2$. As higher polarizabilities are
negligible, the spin-dependent dipole polarizabilities could then be fitted to
data sets which combine polarized and spin-averaged experimental results,
taken at a fixed energy and varying the scattering angle. As starting values 
for the fit,
one might use our $\chi$EFT-results \cite{HGHP03}, as indicated in
Eq.~(\ref{fit}), where we show the spin structure amplitudes up to $l=1$ with
the polarizabilities $\gamma_i(\omega)$ replaced by
$\gamma_i(\omega)+\delta_i$, introducing the fit parameters $\delta_i$. Small 
fit parameters mean correct prediction of the dynamical spin dipole 
polarizabilities within the Small Scale Expansion.
\begin{align}
\begin{split}
  \bar{A}_3^{\mathrm{fit}}(\omega,\,z)
  &=-\frac{4\pi\,W}{M}\,[(\gamma_{E1E1}(\omega)+\delta_{E1E1})
  +z\,(\gamma_{M1M1}(\omega)+\delta_{M1M1})\\
  &+(\gamma_{E1M2}(\omega)+\delta_{E1M2})
  +z\,(\gamma_{M1E2}(\omega)+\delta_{M1E2})]\,\omega^3\nonumber
\end{split}\\
\bar{A}_4^{\mathrm{fit}}(\omega,\,z) &=\frac{4\pi\,
  W}{M}\,\left[-(\gamma_{M1M1}(\omega)+\delta_{M1M1})
  +(\gamma_{M1E2}(\omega)+\delta_{M1E2})\right]\,\omega^3\nonumber\\
\bar{A}_5^{\mathrm{fit}}(\omega,\,z) &=\frac{4\pi\, W}{M}\,
(\gamma_{M1M1}(\omega)+\delta_{M1M1})\,\omega^3\nonumber\\
\bar{A}_6^{\mathrm{fit}}(\omega,\,z) &=\frac{4\pi\,
  W}{M}\,(\gamma_{E1M2}(\omega)+\delta_{E1M2})\,\omega^3
\label{fit}
\end{align}
Thus, one obtains the spin dipole polarizabilities at a definite energy.
Repeating this procedure for various energies gives the energy dependence,
i.e. the dynamics of the $l=1$ spin polarizabilities. Therefore, the
amplitudes Eq.~(\ref{fit}) provide one possible way to extract dynamical spin
polarizabilities directly from the asymmetry observables of the previous
section, using $\chi$EFT.
Note that the $\delta_i$ may show a weak energy dependence.
At first trial, they can be taken 
as energy-independent quantities. This corresponds to a free 
normalization of the spin dipole polarizabilities, assuming the energy 
dependence derived from $\chi$EFT to be correct. This assumption is well 
justified, as at low energies only $\Delta(1232)$ and pion degrees of freedom 
are supposed to give dispersive contributions to the polarizabilities.



\section{Spin Contributions to Spin-Averaged Cross Sections}
\setcounter{equation}{0}
\label{spinav}

\begin{figure}[!htb]
\begin{center}
  \includegraphics*[width=.82\textwidth]{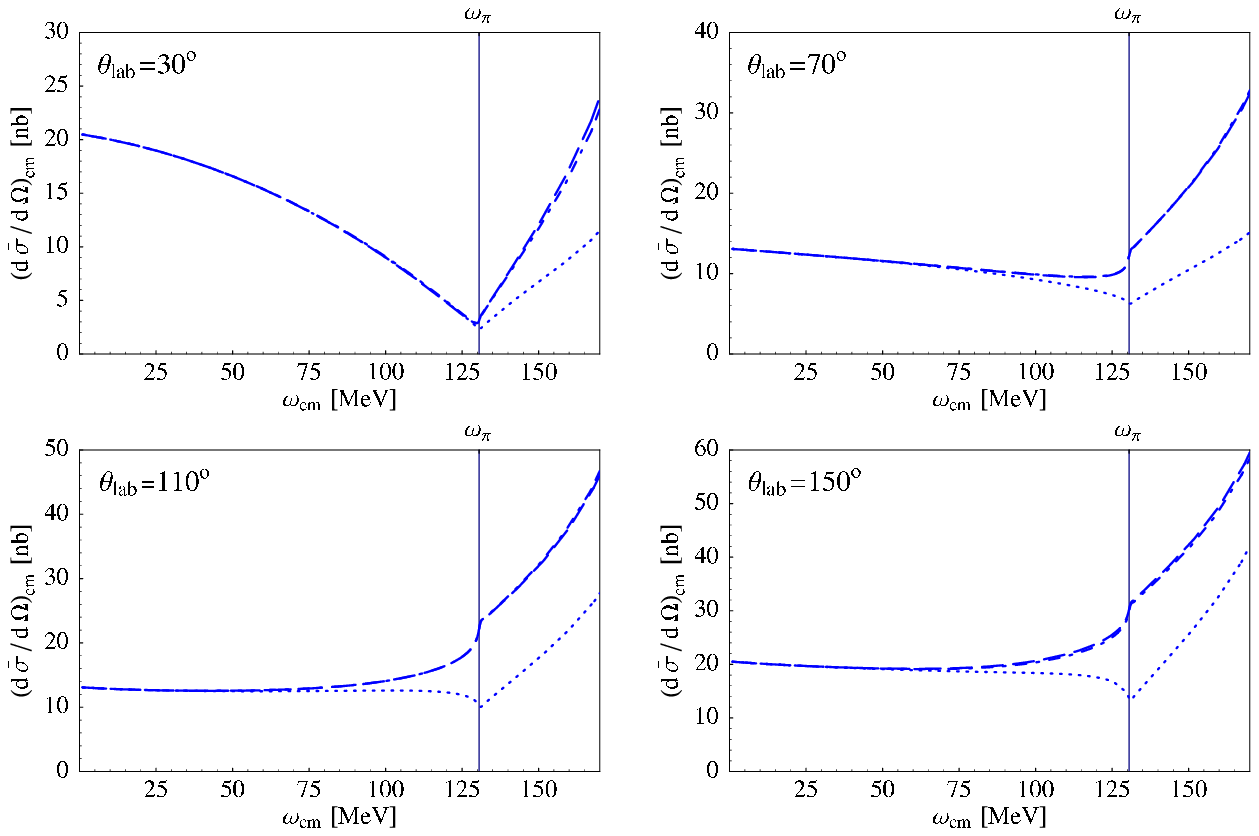}
\caption{Complete
  $\mathcal{O}(\epsilon^3)$-SSE-predictions (dashed) for the spin-averaged
  proton cross section; dotted: spin polarizabilities not included, dotdashed:
  quadrupole polarizabilities not included.}
\label{SSEindiesump}
\end{center}
\end{figure}

\begin{figure}[!htb]
\begin{center}
  \includegraphics*[width=.82\textwidth]{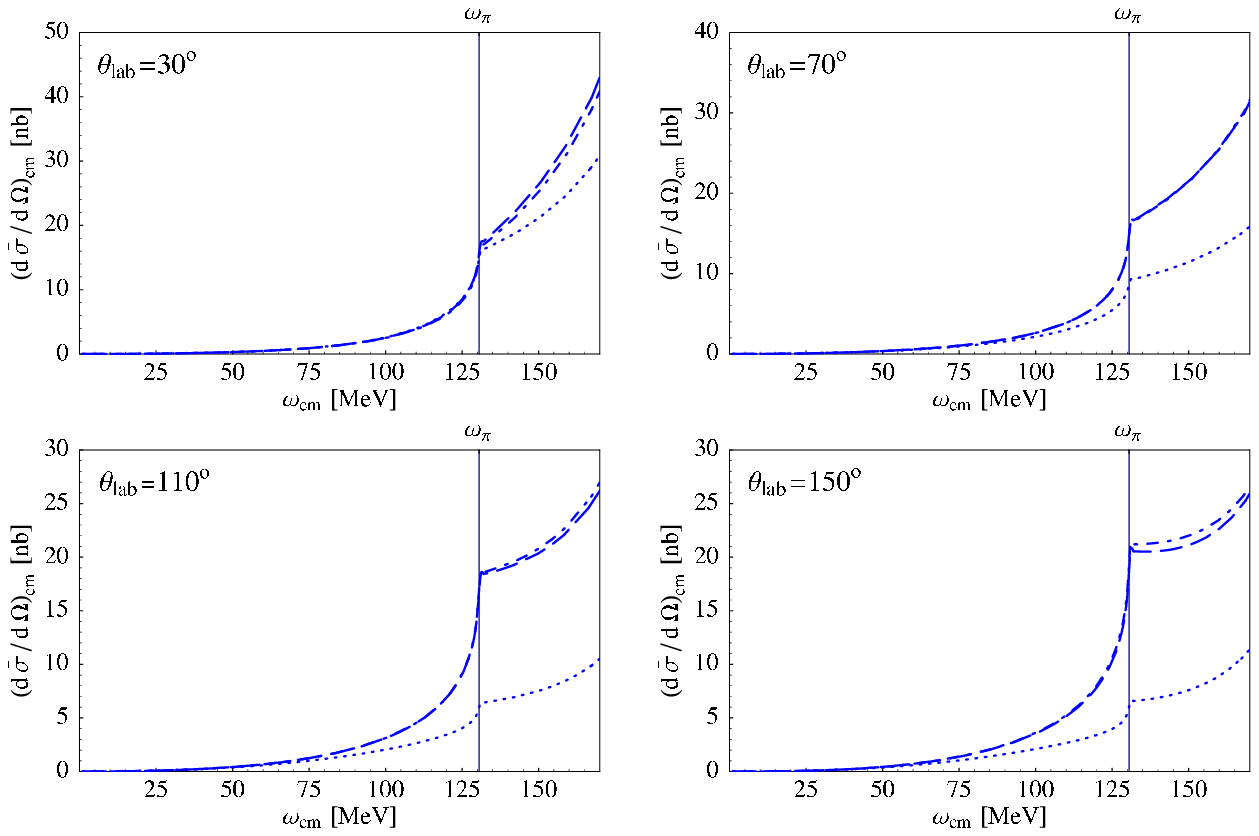}
\caption{Complete
  $\mathcal{O}(\epsilon^3)$-SSE-predictions (dashed) for the spin-averaged
  neutron cross section; dotted: spin polarizabilities not included,
  dotdashed: quadrupole polarizabilities not included.}
\label{SSEindiesumn}
\end{center}
\end{figure}

Before discussing the asymmetries in detail, we briefly turn to the question
which polarizabilities are seen in unpolarized Compton cross sections,
discussing the $\mathcal{O}(\epsilon^3)$ SSE-results partially given already
in~\cite{HGHP03}.
%
As shown in Fig.~\ref{SSEindiesump}, we find a large contribution of the 
dynamical spin polarizabilities to
spin-averaged Compton cross sections on the proton above $\omega\sim
100\;\mathrm{MeV}$.
%
We also show our so far unpublished results for the neutron
(Fig.~\ref{SSEindiesumn}), exhibiting a huge sensitivity on the spin
polarizabilities in the backward direction. This can be well understood, as
the right hand side of Eq.~(\ref{eq:sum}) simplifies to $|A_1|^2+|A_3|^2$ for
$\theta=0^\circ$ and $\theta=180^\circ$. In the forward direction, the
spin-independent amplitude $|A_1|^2$ dominates, in backward direction the
spin-dependent amplitude $|A_3|^2$, as can be seen in App.~\ref{App:A1A3}.

We note also again that any effects of quadrupole polarizabilities are
invisible at the level of the unpolarized cross sections, as has already been
found in \cite{HGHP03} for the proton. It suffices therefore to terminate the
multipole expansion Eq.~(\ref{eq:strucamp}) at the dipole level, which leaves
the six dipole polarizabilities as parameters.

While effects from the spin polarizabilities are non-negligible in unpolarized
experiments, to extract all four of them from such data is clearly illusory.
Thus, double polarized experiments as discussed in the rest of this article
are necessary additional ingredients in a combined multipole analysis.



\section{Proton Asymmetries}
\setcounter{equation}{0}
\label{sec:proton}

We therefore turn now to the results for the asymmetries of the proton.
Analogously to the previous section, we will confirm for each observable that
the quadrupole polarizabilities are negligible. Thus, the multipole expansion
of the amplitude can always be truncated at the dipole level, leaving at most
six parameters. However, it will turn out that not all asymmetries are equally
sensitive on the spin polarizabilities. As expected, most asymmetries are 
indeed governed by the pole part of the amplitudes.

In order to determine which asymmetries are most sensitive to the
structure parts of the Compton amplitudes, and which of the internal low
energy degrees of freedom in the nucleon dominate the structure dependent part
of the cross section, we will first compare three scenarios for each
asymmetry: (i) the result when only the pole terms of the amplitudes are kept;
(ii) the same when the effects from the pion cloud around the nucleon are
added, as described by the leading-one-loop order \HBChiPT result; and 
finally (iii) a
leading-one-loop order calculation in SSE, including also the $\Delta$ as 
dynamical degree of freedom.

An ideal asymmetry should thus fulfill three criteria: It should be large to
give a good experimental signal, it should show sensitivity on the structure
amplitudes, and it should allow a differentiation between the pion cloud and
$\Delta$ resonance contributions in resonant channels, revealing as much as 
possible about the role
of at least these low energy degrees of freedom in the nucleon. In
Sect.~\ref{sec:neutron}, we will repeat this presentation for the neutron
asymmetries.  To simplify connection to experiment, we give the scattering
angle in the following plots in the lab-frame.

Similar plots for the nucleon asymmetries are already shown in \cite{Babusci},
using Dispersion Theory techniques. Direct comparison to those plots is
however not possible because of a different choice of angles -- the authors of
\cite{Babusci} concentrated on the extreme angles $0^\circ$, $90^\circ$,
$180^\circ$, whereas we cover the whole experimentally accessible angular
spectrum. Nevertheless, qualitative agreement between our $\chi$EFT results
and \cite{Babusci} can be deduced.

We emphasize also that our predictions are parameter-free, as all constants
are determined from unpolarized Compton scattering on the proton,
\cite{HGHP03}. In the following the fit parameters $\delta_i$ introduced in 
Eq.~(\ref{fit}) are
all set to zero, as no measured asymmetries exist at this point.

\subsection{Nucleon Spin Parallel Photon Momentum}
\label{sec:pparanti}

\subsubsection{Comparison: Pole-, \HBChiPT- and SSE-calculation of 
  $\Sigma_z^p$}

\begin{figure}[!htb]
\begin{center}
  \includegraphics*[width=.82\textwidth]{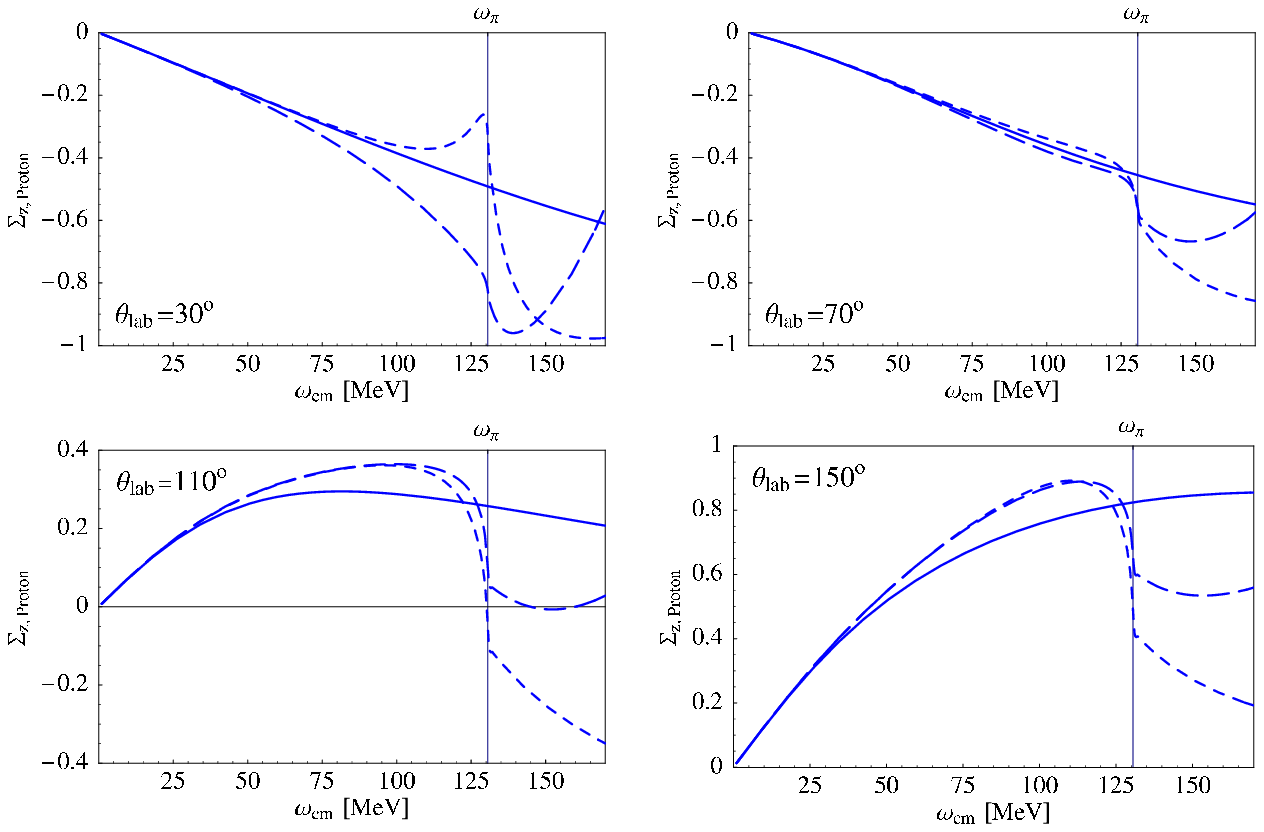}
\caption{
  $\mathcal{O}(p^3)$-\HBChiPT- (shortdashed) and
  $\mathcal{O}(\epsilon^3)$-SSE-predictions (longdashed) for the proton
  asymmetry $\Sigma_z^p$;
  the solid line describes the third order pole contributions.}
\label{bornasyp}
\end{center}
\end{figure}

As one can see in Fig.~\ref{bornasyp}, the proton asymmetry $\Sigma_z^p$
reaches values of $\mathcal{O}(1)$ and is therefore quite large, although it
vanishes for $\omega=0$, due to the vanishing difference and the finite static
spin-averaged cross section, given by the familiar Thomson-limit.  This is
valid independently of the scattering angle.

Comparing the three curves in Fig.~\ref{bornasyp} -- third order pole,
$\mathcal{O}(p^3)$-\HBChiPT and $\mathcal{O}(\epsilon^3)$-SSE -- one
recognizes the strong influence of the pole amplitudes, given by
Eq.~(\ref{eq:pole}). This is exactly what one expects for the charged proton,
and can also be deduced from Eqs.~(\ref{eq:strucamp}, \ref{eq:pole},
\ref{eq:paranti}): The asymmetry starts linearly in $\omega$, while the
leading term of the structure part of $\Sigma_z^p$ is proportional to
$\omega^3$, as there is no term in Eq.~(\ref{eq:paranti}) that contains only
spin-independent amplitudes. As we are interested in information about the
structure of the nucleon, i.e.~in the deviation of the dashed lines from the
solid (pole contributions only) line in Fig.~\ref{bornasyp}, and as this
deviation is not as strong as later in $\Sigma_x^p$ and in the neutron
asymmetries, $\Sigma_z^p$ does not seem to be the ideal choice among the
considered quantities to examine the nucleon structure.

Concerning the explicit $\Delta$ degrees of freedom, we see sizeable
contributions only above $\omega_\pi$. The only exception is noticed in
extreme forward direction, but this is an artifact of the asymmetry, which is
extremely sensitive at small angles due to the small spin-averaged cross
section at $\omega_\pi$ (Fig.~\ref{SSEindiesump}), and neither visible in the 
difference $\mathcal{D}_z$ nor in the spin-averaged cross section.

The structure of $\Sigma_z^p$ varies a lot for the different angles. It is
negative in forward and positive in backward direction.  This can be explained
-- at least for low energies ($\le 120$ MeV) -- looking at the amplitudes
$A_1$ and $A_3$ in Fig.~\ref{A1A3} because for $\theta=0^\circ$ and
$\theta=180^\circ$ the right hand side of Eq.~(\ref{eq:paranti}) reduces to
$-2\,A_1\,A_3$.
The proton amplitude $A_3$ starts with a falling slope in forward and with a
rising slope in backward direction, while $A_1$ is negative below the pion
production threshold for all angles under consideration. The spin-averaged
cross section is positive for all angles and energies by definition.

At $\omega_\pi\approx 131$ MeV in the cm system, the cusp at the pion
production threshold is clearly visible for most of the angles.  This cusp
arises since the amplitudes become complex at the threshold.  Polarized cross
sections are much more sensitive on the fine structure of the nucleon than
their unpolarized pendants.
Therefore, our results might considerably deviate from experiment above
threshold, due to sizeable uncertainties in our imaginary parts. Nevertheless,
qualitative agreement should be fulfilled, so we use the same plot 
range as for
the unpolarized results in \cite{HGHP03}, with a maximum photon energy of
$170$ MeV.  In \cite{Babusci}, the plots end below threshold since its low
energy expansion of the polarizabilities cannot reproduce the non-analyticity
of the pion production threshold.


%

\subsubsection{Spin and Quadrupole Contributions to $\Sigma_z^p$}

\begin{figure}[!htb]
\begin{center}
  \includegraphics*[width=.82\textwidth]{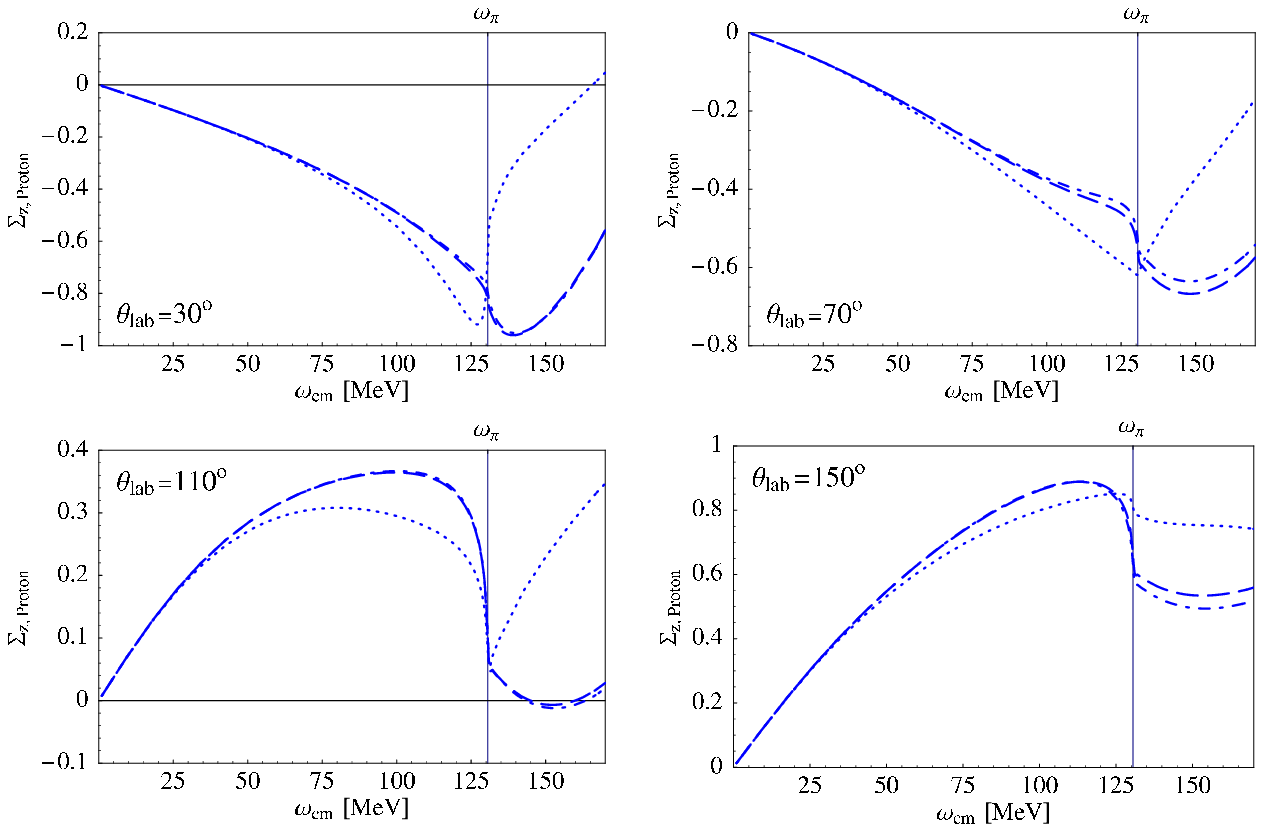}
\caption{
  Complete $\mathcal{O}(\epsilon^3)$-SSE-predictions (dashed) for the proton
  asymmetry $\Sigma_z^p$; dotted: spin polarizabilities not included,
  dotdashed: quadrupole polarizabilities not included.}
\label{SSEindieasyp}
\end{center}
\end{figure}
\begin{figure}[!htb]
\begin{center}
  \includegraphics*[width=.82\textwidth]{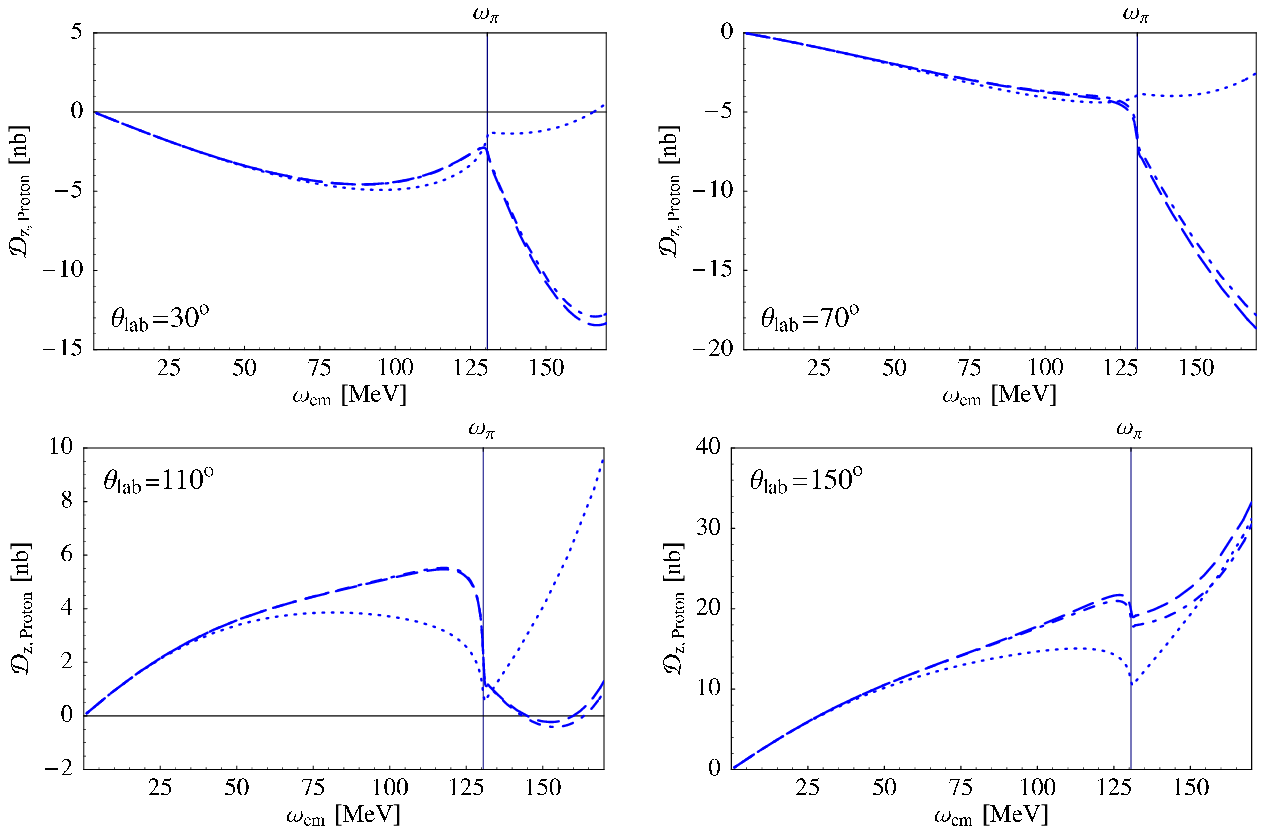}
\caption{
  Complete $\mathcal{O}(\epsilon^3)$-SSE-predictions (dashed) for the proton
  difference $\mathcal{D}_z^p$; dotted: spin polarizabilities not included,
  dotdashed: quadrupole polarizabilities not included.}
\label{SSEindiediffp}
\end{center}
\end{figure}

%

The asymmetry (Fig.~\ref{SSEindieasyp}), and especially the difference (Fig.
\ref{SSEindiediffp}), exhibit only a weak dependence on the spin
polarizabilities in forward direction below $\omega_\pi$, which is no
surprise, as in Fig.~\ref{A1A3} there are nearly no structure contributions to
$A_3^p$ visible below 100 MeV.
In the backward direction, the sensitivity on the $\gamma_i$'s is large,
especially in the difference.

The sharp rise of the result without spin polarizabilities in
Fig.~\ref{SSEindieasyp} above the pion production threshold in forward
direction is due to the sharply rising difference and the small spin-averaged
cross section which enters the denominator presented in
Fig.~\ref{SSEindiesump}.

In the literature, e.g.~in \cite{Sandorfi}, the pion pole
(Fig~\ref{poldiagramme}.d) is often considered as one of the structure
diagrams, giving a contribution to the static backward spin polarizability
$\gamma_\pi$, which is much larger than all the other contributions.  We treat
the term as pole, as it contains a pion pole in the $t$-channel. So the
question arises why we are sensitive to the spin polarizabilities, despite of
having removed this supposedly dominant part from them. The reason is that
the pion pole dominates over the structure part of $\gamma_\pi$ only for low
energies. The pion pole contribution to $\gamma_\pi(\omega)$ describes a
Lorentzian (Eq.~(\ref{eq:pole})) and becomes smaller than the structure
contribution above 100 MeV, as the latter one rises due to the increasing
values of $\gamma_{E1E1}(\omega)$ and $\gamma_{M1M1}(\omega)$ \cite{HGHP03}.




It is crucial to notice that the quadrupole polarizabilities ($l=2$) play
again a negligibly small role, see
Figs.~\ref{SSEindieasyp} and~\ref{SSEindiediffp}. The most important
quadrupole contribution is observed at $70^\circ$ and $150^\circ$, but the
relative size is still $<0.1$ and therefore presumably within the experimental
error bars. As repeatedly stated, that these contributions are small is
mandatory if one wants to determine spin polarizabilities via polarized cross
section data, because only then can the multipole expansion be truncated at 
$l=1$ as in Eqs.~(\ref{eq:strucamp}, \ref{fit}).



\subsection{Nucleon Spin Perpendicular to Photon Momentum}

\subsubsection{Comparison: Pole-, \HBChiPT- and SSE-calculation of 
  $\Sigma_x^p$}

\begin{figure}[!htb]
\begin{center}
  \includegraphics*[width=.82\textwidth] {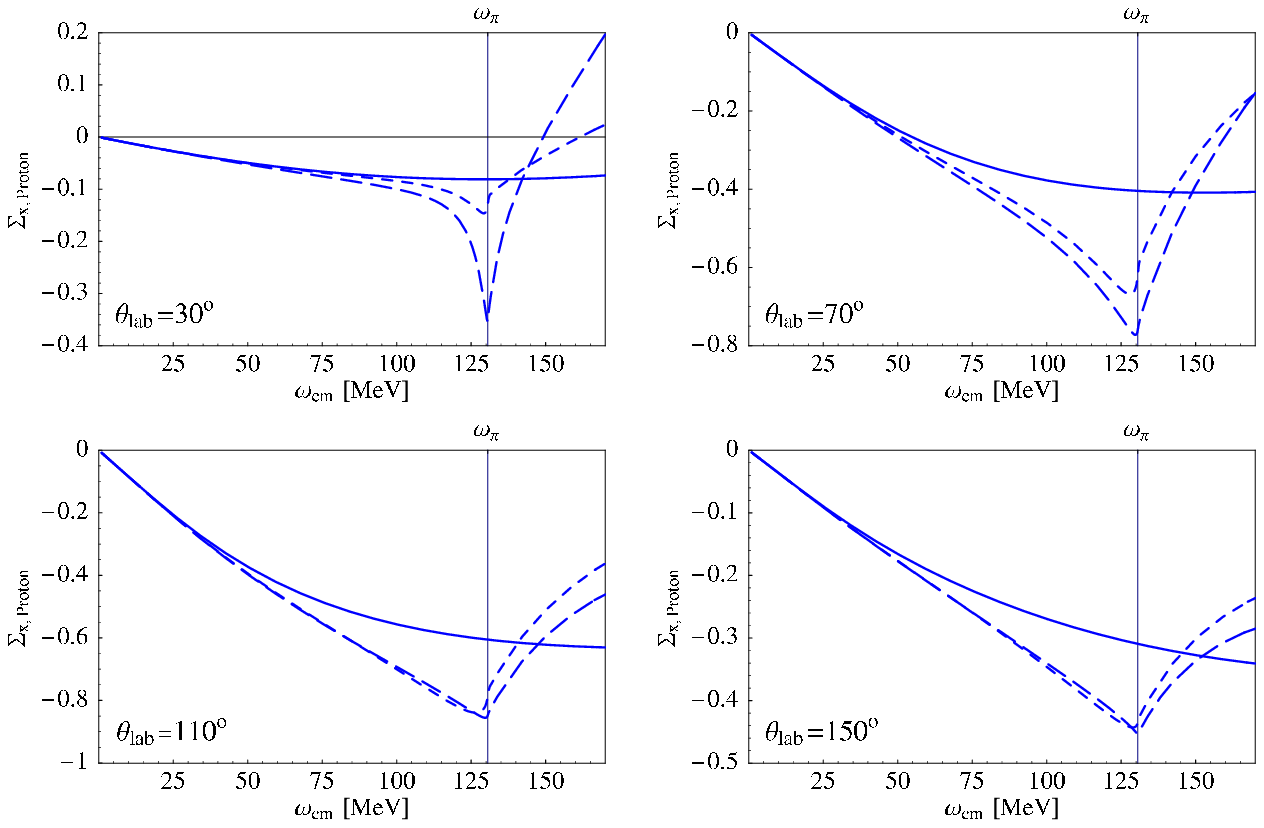}
\caption{
  $\mathcal{O}(p^3)$-\HBChiPT- (shortdashed) and
  $\mathcal{O}(\epsilon^3)$-SSE-predictions (longdashed) for the proton
  asymmetry $\Sigma_x^p$;
  the solid line describes the third order pole contributions.}
\label{bornvertasyp}
\end{center}
\end{figure}

The asymmetry $\Sigma_x^p$
in Fig.~\ref{bornvertasyp} looks quite similar for the different angles: It
always starts with a falling slope and exhibits a sharp minimum at the pion
cusp, therefore staying negative in a wide energy range.  This behavior is no
surprise, as the leading term in Eq.~(\ref{eq:rightleft}) for the proton for
$\theta\approx0^\circ,\;\theta\approx180^\circ$, i.e.~$\sin^2\theta\approx 0$,
is $A_3\,(A_3-A_1)\,\sin\theta\,\cos\theta$, which is the only term including
$A_1$ and therefore dominating for low energies, as $A_1^p$ contains the
Thomson-limit. In both forward and backward direction $A_3-A_1>0$, and
$A_3<(>)\;0$ for small (large) angles and low energies. The factor
$\cos\theta$ gives an additional minus sign in backward direction.

Even more striking than for $\Sigma_z^p$ is the weak sensitivity of the
asymmetry $\Sigma_x^p$ on explicit $\Delta$ degrees of freedom.  Once again,
the only exception to this rule is the extreme forward direction because of
the small spin-averaged cross section which enhances the small deviations
between the HB$\chi$PT- and the SSE-calculation of the difference
$\mathcal{D}_x^p$ and makes $\Sigma_x^p$ extremely sensitive on errors.
Therefore, we consider the forward angle regime as inconvenient for measuring
proton asymmetries. In the other panels of Fig.~\ref{bornvertasyp} the
$\Delta$-dependence cancels in the asymmetry, whereas we found the
$\Delta(1232)$ resonance to give sizeable contributions to both the difference
and the sum. This is one example that an asymmetry actually hides interesting
physical information.

The dominance of the pole amplitudes is -- as in $\Sigma_z^p$ -- clearly
visible. The argument is the same as the one given in Sect.
\ref{sec:pparanti}. Nonetheless, we find $\Sigma_x^p$ more sensitive on the
nucleon structure than $\Sigma_z^p$, especially around $\omega_\pi$.

%

\subsubsection{Spin and Quadrupole Contributions to $\Sigma_x^p$}

\begin{figure}[!htb]
\begin{center}
  \includegraphics*[width=.82\textwidth] {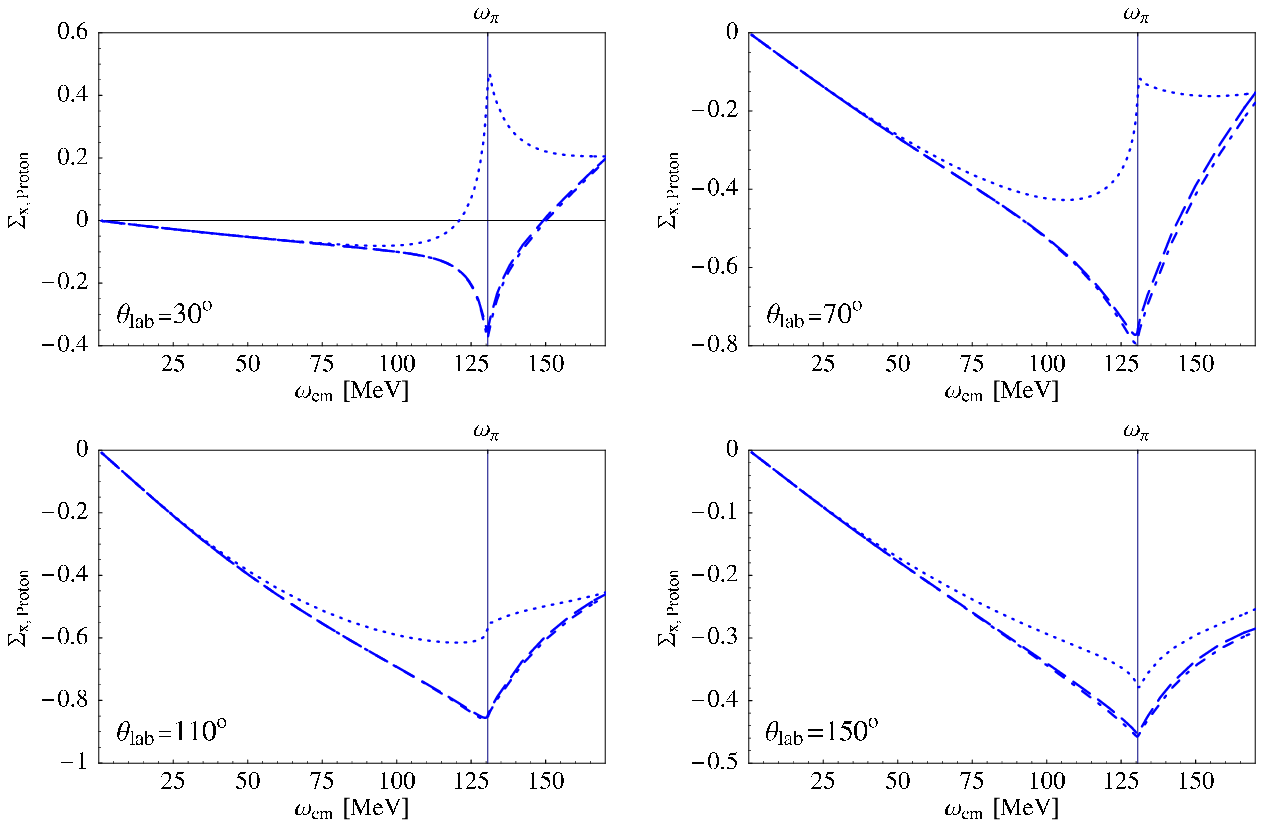}
\caption{ Complete
  $\mathcal{O}(\epsilon^3)$-SSE-predictions (dashed) for the proton asymmetry
  $\Sigma_x^p$; dotted: spin polarizabilities not included, dotdashed:
  quadrupole polarizabilities not included.}
\label{SSEindievertasyp}
\end{center}
\end{figure}
\begin{figure}[!htb]
\begin{center}
  \includegraphics*[width=.82\textwidth]{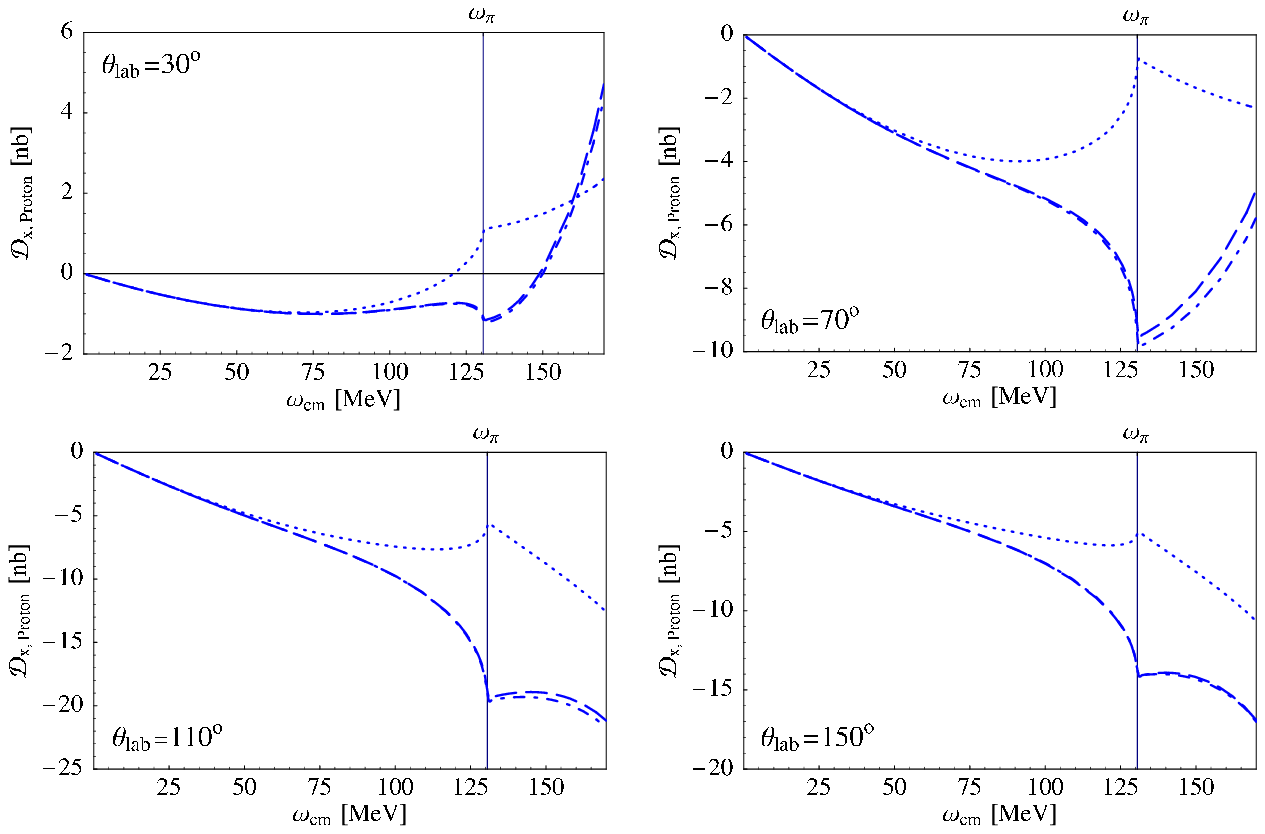}
\caption{
  Complete $\mathcal{O}(\epsilon^3)$-SSE-predictions (dashed) for the proton
  difference $\mathcal{D}_x^p$;
  dotted: spin polarizabilities not included, dotdashed: quadrupole
  polarizabilities not included.}
\label{SSEindievertdiffp}
\end{center}
\end{figure}


As one can see in Figs.~\ref{SSEindievertasyp} and~\ref{SSEindievertdiffp}, 
$\Sigma_x^p$ and $\mathcal{D}_x^p$ are very sensitive
on the spin polarizabilities for all angles. Therefore -- and because of our
findings in the previous subsection -- this configuration (nucleon spin
perpendicular to the photon momentum) seems to be more convenient than spin
parallel photon momentum, to examine the spin structure of the nucleon. In the
backward direction, the spin dependence of the asymmetry is less pronounced
than in forward direction.



The quadrupole contributions are extremely small.


\section{Neutron Asymmetries}
\setcounter{equation}{0}
\label{sec:neutron}

 In the absence of stable
single neutron targets, the following results for the neutron have to be
corrected for binding and meson exchange effects inside light nuclei, a task
which will be the scope of future work. Here, we present the neutron results
to guide considerations on future experiments using polarized deuterium or
$\,^3\mathrm{He}$, e.g.~\cite{Gao}.

As in the proton case, the neutron asymmetries reach quite large values of
$\mathcal{O}(1)$ as the photon energy increases. In the neutron, pole
contributions might be expected to be small, because it is uncharged and thus
only the pion pole and anomalous magnetic moment contribute. On the other
hand, spin polarizabilities are then not enhanced by interference with large
pole amplitudes. Therefore, whether and which neutron asymmetries are
sensitive to the structure parts, and hence to the $\gamma_i$'s, must be
investigated carefully.

We follow the same line of presentation as outlined at the beginning of
Sect.~\ref{sec:proton} for the proton asymmetries: First, we investigate which
internal degrees of freedom are seen in a specific asymmetry, and then show
that quadrupole polarizabilities give negligible contributions. Thus, the
asymmetries most sensitive to spin polarizabilities are identified.

\subsection{Nucleon Spin Parallel Photon Momentum}

\subsubsection{Comparison: Pole-, \HBChiPT- and SSE-calculation of 
  $\Sigma_z^n$}

\begin{figure}[!htb]
\begin{center}
  \includegraphics*[width=.82\textwidth] {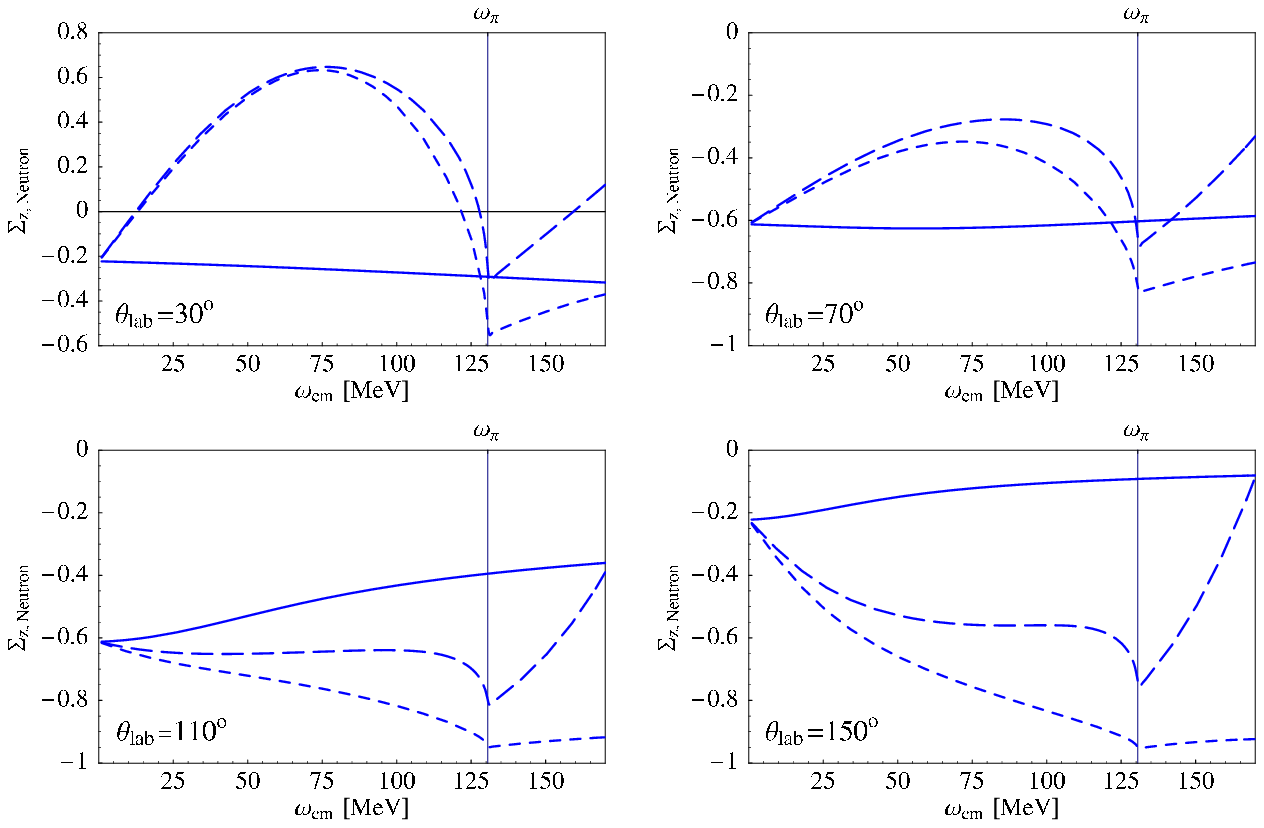}
\caption{
  $\mathcal{O}(p^3)$-\HBChiPT- (shortdashed) and
  $\mathcal{O}(\epsilon^3)$-SSE-predictions (longdashed) for the neutron
  asymmetry $\Sigma_z^n$;
  the solid line describes the third order pole contributions.}
\label{bornasyn}
\end{center}
\end{figure}

Comparing Fig.~\ref{bornasyn} to the proton analogs, Figs. \ref{bornasyp} and
\ref{bornvertasyp}, we notice that the neutron is much more sensitive on the
structure amplitudes. The pole curves show only a weak energy-dependence,
so that nearly the whole dynamics is given by the neutron polarizabilities.
This minor influence of the pole amplitudes is due to the vanishing third
order pole contributions to $A_1$ and $A_2$, which make the difference
Eq.~(\ref{eq:paranti}) start with a term proportional to $\omega^2$, whereas
the leading structure part is $\mathcal{O}(\omega^3)$. The spin-averaged
cross section starts with $\omega^2$, rendering finite static values of
$\Sigma_z^n$. The angular dependence of this static value can be derived from
Eqs.~(\ref{eq:Bornn}, \ref{eq:paranti}, \ref{eq:sum}) as
\begin{equation}
\Sigma_z^n(\omega=0,\;\theta)=\frac{4\,\sin^2\theta}{-5+\cos(2\,\theta)}\;\;.
\end{equation}
The structure sensitivity of the neutron is also visible in the huge
sensitivity of $\Sigma_z^n$ to the $\Delta$ resonance which influences the
polarized cross sections considerably even for very low energies. As is
well-known, the influence of the $\Delta(1232)$ increases with increasing
angle.

Concerning the shape of the asymmetry, one recognizes a similar behaviour for
the whole angular spectrum. It always reaches a local minimum at the pion
cusp. A precise interpretation of the shape of $\Sigma_z^n$ is hard to give,
as the denominator has the leading power $\omega^2$, while it was $\omega^0$
in the proton case. Therefore, it is more instructive to look at the
difference (Fig.~\ref{SSEindiediffn}). For small angles this quantity first
rises, while in backward direction it becomes negative from the very
beginning. The reason is perfectly clear from the amplitudes $A_1$ and $A_3$
in App. \ref{App:A1A3}, as the product $A_1\cdot A_3$ is negative for low
energies in forward direction, positive in backward direction. Recall that for
the neutron, $A_1=\bar{A}_1$ contains only structure contributions,
Eq.~\ref{eq:pole}.



\subsubsection{Spin and Quadrupole Contributions to $\Sigma_z^n$}

\begin{figure}[!htb]
\begin{center}
  \includegraphics*[width=.82\textwidth]{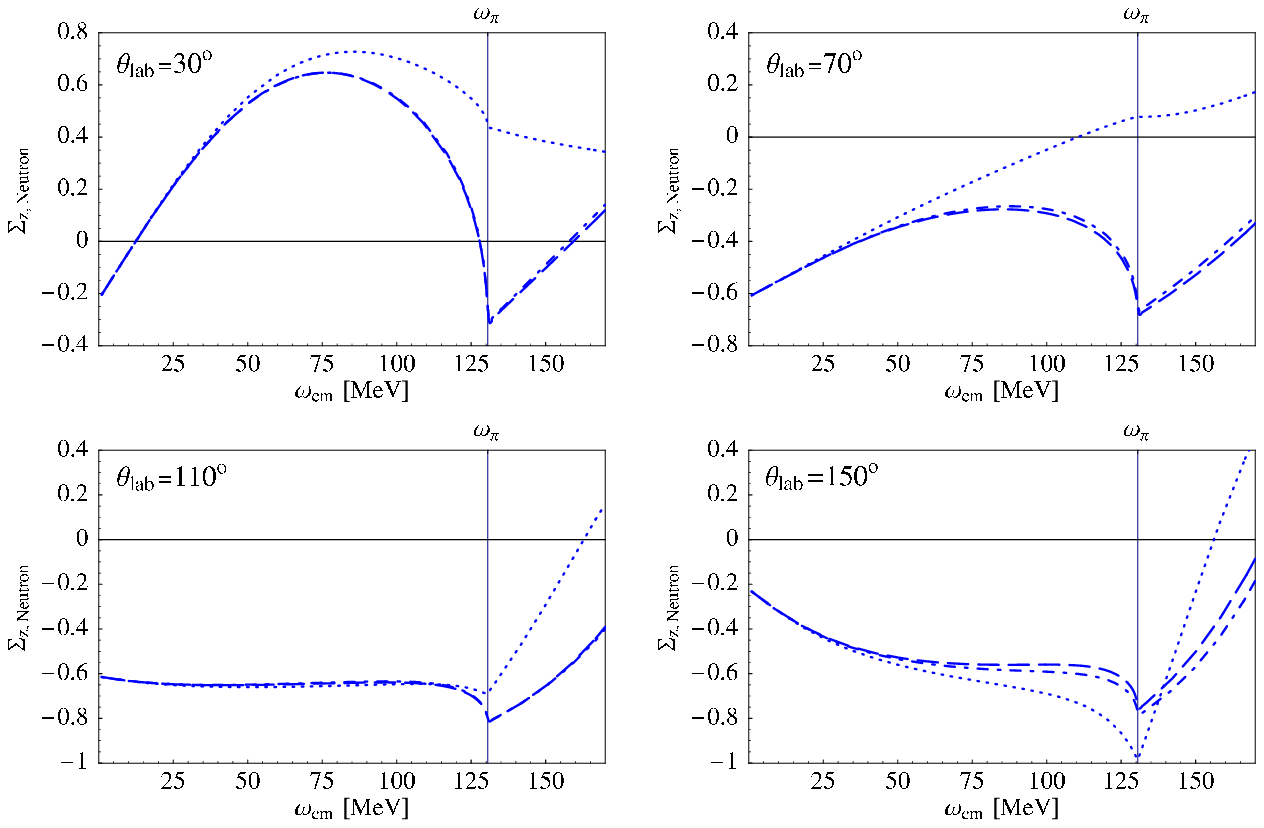}
\caption{
  Complete $\mathcal{O}(\epsilon^3)$-SSE-predictions (dashed) for the neutron
  asymmetry $\Sigma_z^n$; dotted: spin polarizabilities not included,
  dotdashed: quadrupole polarizabilities not included.}
\label{SSEindieasyn}
\end{center}
\end{figure}

\begin{figure}[!htb]
\begin{center}
  \includegraphics*[width=.82\textwidth]{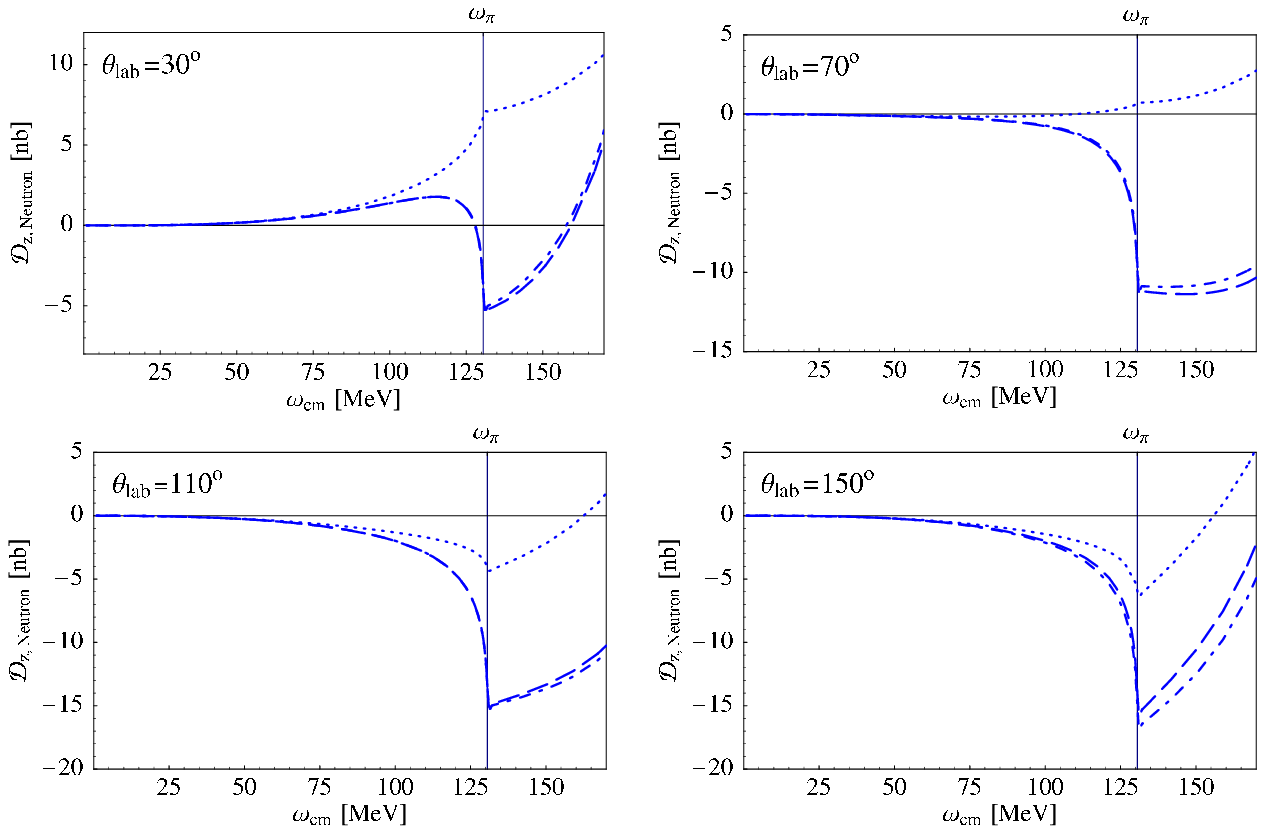}
\caption{
  Complete $\mathcal{O}(\epsilon^3)$-SSE-predictions (dashed) for the neutron
  difference $\mathcal{D}_z^n$;
  dotted: spin polarizabilities not included, dotdashed: quadrupole
  polarizabilities not included.}
\label{SSEindiediffn}
\end{center}
\end{figure}

Fig.~\ref{SSEindieasyn} exhibits that there are sizeable spin contributions to
the asymmetry $\Sigma_z^n$ for each angle, but for $\theta=110^\circ$ they
nearly completely vanish below the pion production threshold. The reason is
cancellation in the division of the difference (Fig.~\ref{SSEindiediffn}) by
the sum (Fig.~\ref{SSEindiesumn}). This cancellation arises as the shape of
the three curves in Fig.~\ref{SSEindiediffn} for $\theta=110^\circ$ is
nearly exactly symmetric to the three spin-averaged cross section curves at
the same angle with respect to the $\omega$-axis. Therefore, dividing both
results by each other hides the spin contribution.
In the difference, one recognizes a decreasing spin dependence with increasing
angle, which can again be explained by $A_1$ and $A_3$.  In forward direction,
$A_3^\mathrm{pole}$ starts with a falling slope and stays negative for the
energy range we are considering. Adding the structure part of $A_3$,
i.e.~including the spin polarizabilities, the amplitude changes sign roughly
at the pion mass.  Therefore we see a completely different behavior of
$A_1\cdot A_3^\mathrm{pole}$ and $A_1\cdot A_3$ in forward direction. In
backward direction $A_1,\;A_3$ and $A_3^\mathrm{pole}$ are all positive below
$\omega_\pi$, resulting in very similar curves.



As in the proton case we find the quadrupole part to be negligibly small
within the accuracy of this analysis (Fig.~\ref{SSEindieasyn}).


\subsection{Nucleon Spin Perpendicular to Photon Momentum}

\subsubsection{Comparison: Pole-, \HBChiPT- and SSE-calculation of 
  $\Sigma_x^n$}

\begin{figure}[!htb]
\begin{center}
  \includegraphics*[width=.82\textwidth] {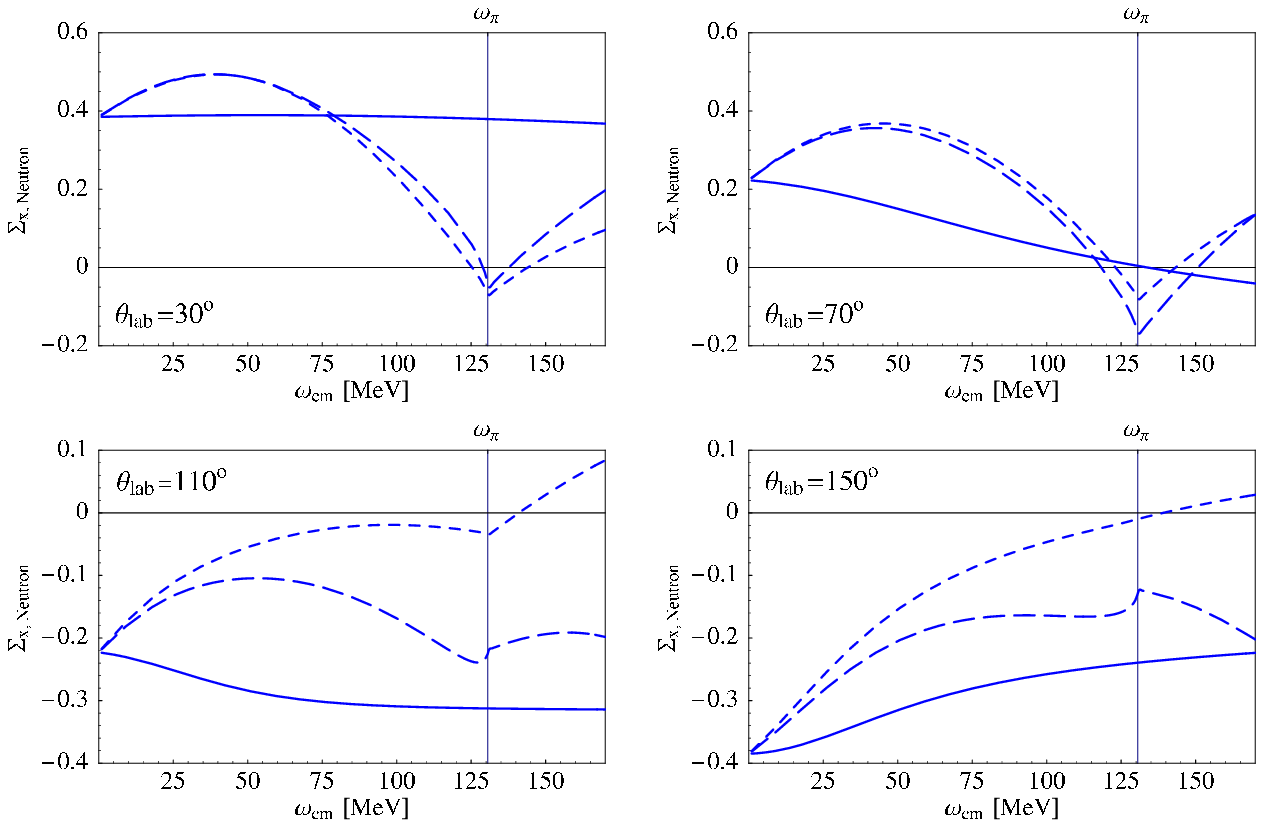}
\caption{
  $\mathcal{O}(p^3)$-\HBChiPT- (shortdashed) and
  $\mathcal{O}(\epsilon^3)$-SSE-predictions (longdashed) for the neutron
  asymmetry $\Sigma_x^n$;
  the solid line describes the third order pole contributions.}
\label{bornvertasyn}
\end{center}
\end{figure}

The shape of the asymmetry $\Sigma_x^n$
in Fig.~\ref{bornvertasyn}
with the minimum at $\omega_\pi$ is similar to $\Sigma_z^n$
(Fig.~\ref{bornasyn}), at least in forward direction. The curve is shifted
downward with increasing angle $\theta$. An explanation for this behaviour can
be given, though it is not obvious, as for
$\theta\approx0^\circ,\;\theta\approx180^\circ$ there remain five terms in
Eq.~(\ref{eq:rightleft}):
$[A_3\,(A_3-A_1)\,\cos\theta+A_3\,A_4\,(1+\cos^2\theta)
+A_3\,A_5\,(3\,\cos^2\theta-1)+2\,A_3\,A_6\,\cos\theta]\,\sin\theta\approx
[A_3\,(A_3-A_1)\,\cos\theta+2\,A_3\,(A_4+A_5+A_6\,\cos\theta)]\,\sin\theta$.
As can be read off Eq.~(\ref{eq:Bornn}), the leading pole terms -- which are
linear in $\omega$ -- vanish in the sum $A_4+A_5+A_6\,\cos\theta$. Therefore
the lowest order in $\omega$ of $A_3\,(A_4+A_5+A_6\,\cos\theta)$ is
$\omega^4$, since the spin-dependent structure amplitudes start with
$\omega^3$.  This is two orders in $\omega$ above the leading order of
$A_3\,(A_3-A_1)\,\cos\theta$, which therefore is the leading term for small
energies -- at least for $|\cos\theta|\gg 0$.  As discussed before, the
product $A_1\,A_3$ is negative for low energies in forward direction, leading
to a positive slope of the difference and therefore to positive values for the
asymmetry. In backward direction, $(A_3-A_1)$, as well as $A_3$, is positive,
which gives a negative asymmetry as $\cos\theta<0$. The angular dependence of
the static value is determined by the pole contributions. It is
\begin{equation}
\Sigma_x^n(\omega=0,\;\theta)=\frac{4\,\sin\theta\,\cos\theta}
                                   {5-\cos(2\,\theta)}\;\;,
\end{equation}
but as for $\Sigma_z^n$, the dynamics of $\Sigma_x^n$ is completely dominated
by the neutron polarizabilities.

Another interesting feature in Fig.~\ref{bornvertasyn} is the fact, that the
explicit $\Delta$ degrees of freedom only play a minor role in forward
direction but dominate for large angles.

%

\subsubsection{Spin and Quadrupole Contributions to $\Sigma_x^n$}

\begin{figure}[!htb]
\begin{center}
  \includegraphics*[width=.82\textwidth] {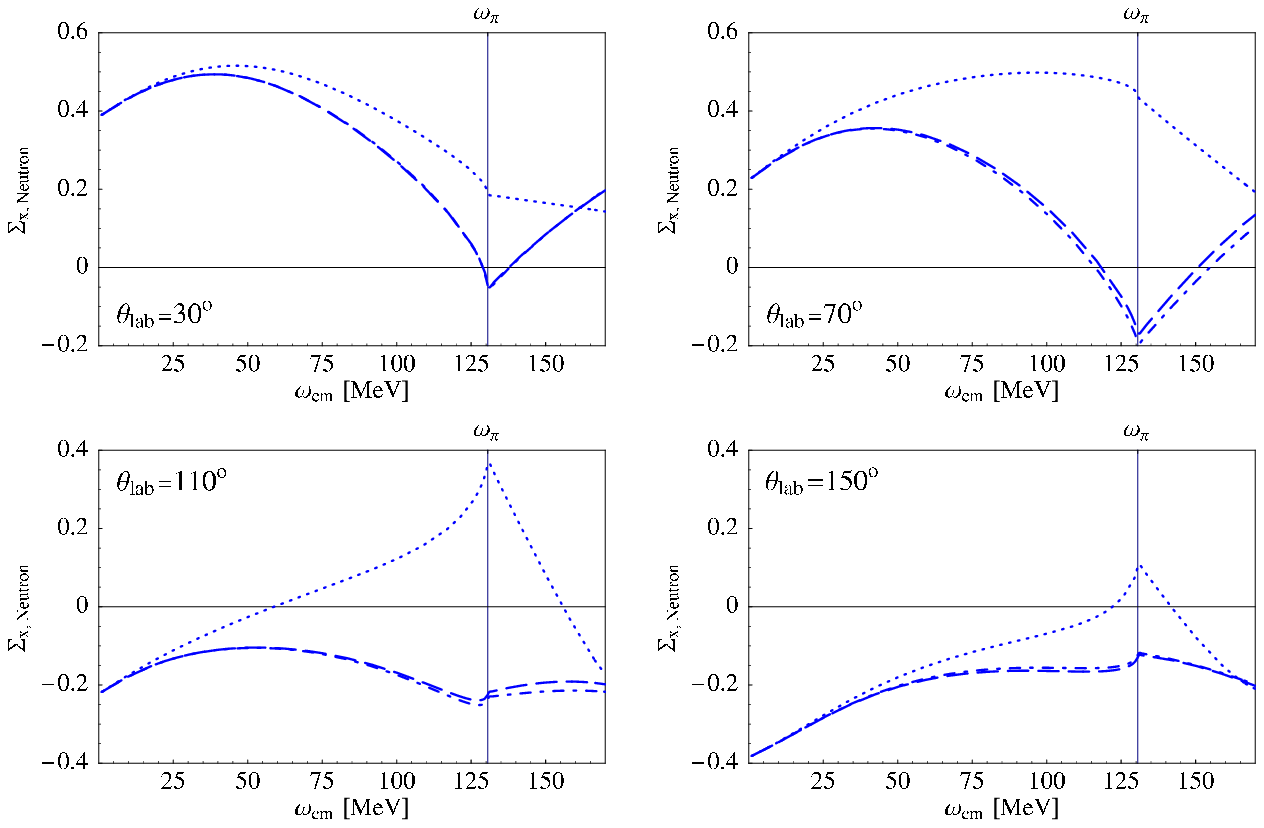}
\caption{
  Complete $\mathcal{O}(\epsilon^3)$-SSE-predictions (dashed) for the neutron
  asymmetry $\Sigma_x^n$; dotted: spin polarizabilities not included,
  dotdashed: quadrupole polarizabilities not included.}
\label{SSEindievertasyn}
\end{center}
\end{figure}
\begin{figure}[!htb]
\begin{center}
  \includegraphics*[width=.82\textwidth]{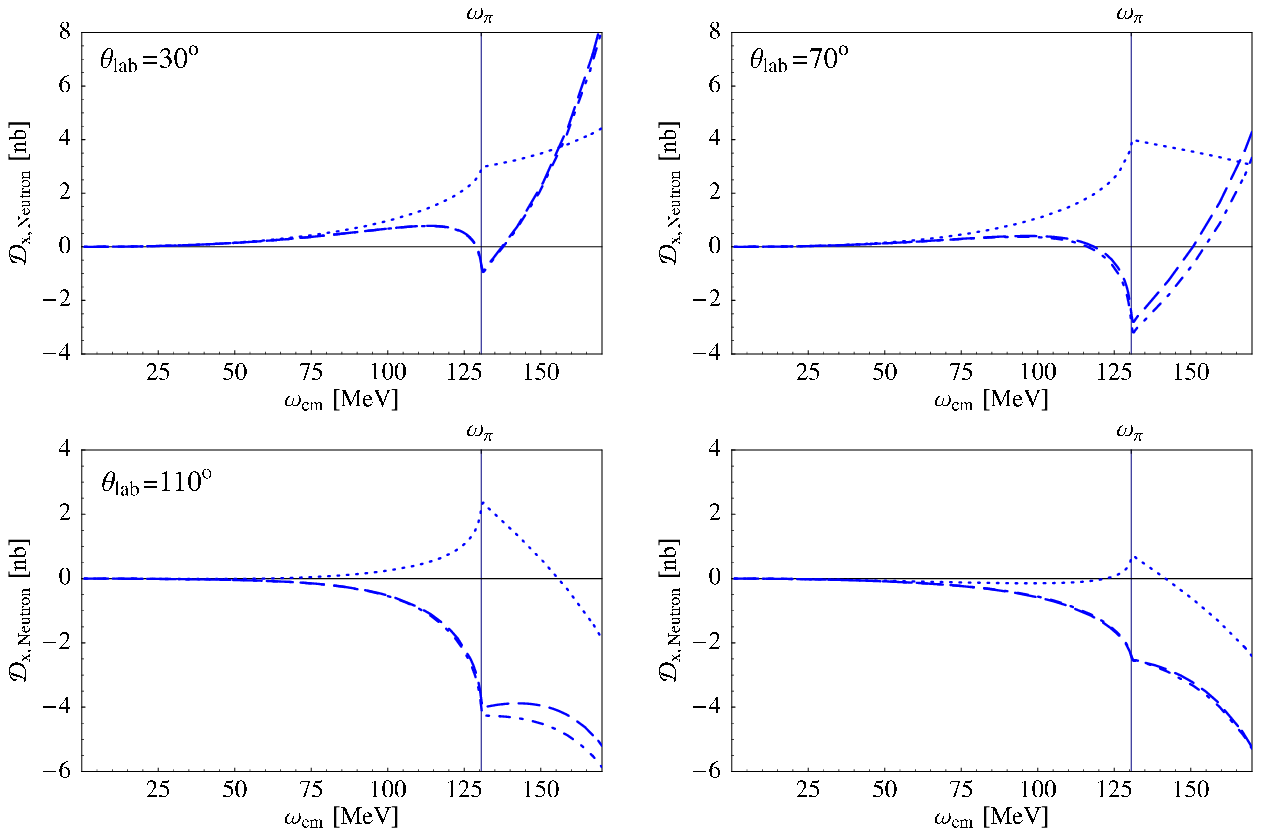}
\caption{
  Complete $\mathcal{O}(\epsilon^3)$-SSE-predictions (dashed) for the neutron
  difference $\mathcal{D}_x^n$;
  dotted: spin polarizabilities not included, dotdashed: quadrupole
  polarizabilities not included.}
\label{SSEindievertdiffn}
\end{center}
\end{figure}

Turning to Figs.~\ref{SSEindievertasyn} and
\ref{SSEindievertdiffn}, $\Sigma_x^n$ exhibits of all asymmmetries by far 
the largest
sensitivity on the spin polarizabilities.
Therefore, an
experiment with the nucleon spin aligned perpendicular to the photon momentum
seems from the theorist's point of view to be the most promising of the
considered configurations to extract the 
spin polarizabilities. The weakest dependence on the $\gamma_i$'s at low
energies of $\Sigma_x^n$ occurs in extreme forward direction. This again can
be explained looking at $A_1$ and $A_3$ (Fig.~\ref{A1A3}): In the backward
direction, $A_3^\mathrm{pole}<A_3$ and therefore also
$(A_3^\mathrm{pole}-A_1)<(A_3-A_1)$, giving a much larger absolute value when
spins are included. In the forward direction, $A_3$ differs weakly from
$A_3^\mathrm{pole}$ for $\omega\le 110$ MeV, so that spin polarizability
effects appear only for higher energies.



As in $\Sigma_z^n$, the quadrupole polarizabilities in $\Sigma_x^n$ are
negligibly small (Fig.~\ref{SSEindievertasyn}). One observes the strongest
contributions around $\theta=90^\circ$; a simple answer to this phenomenon
cannot be given, albeit it is clear that $\Sigma_x^n$ should be most sensitive
at a scattering angle around $90^\circ$, as the overall factor $\sin\theta$
reaches its maximum (Eq.~(\ref{eq:rightleft})), but this is a general feature
and not only concerning the quadrupole polarizabilities.

\absatz
So as a short conclusion of Sects. \ref{sec:proton} and \ref{sec:neutron} we
find a much stronger sensitivity of the neutron asymmetries on the nucleon 
structure, while the
proton asymmetries are dominated by pole terms up to at least $50$ MeV. 
Contributions of the $\Delta(1232)$ resonance are crucial only for certain 
asymmetries and angles.
For both nucleons,
the spin configuration $\Sigma_x$ turned out to be more sensitive on the
nucleon spin structure than $\Sigma_z$. Dynamical quadrupole contributions are
negligible in each of the considered cases.


\section{Conclusion}
\setcounter{equation}{0}
\label{sec:con}

In this work, we examined spin-averaged and double polarized nucleon Compton
cross sections in the framework of Chiral Effective Field Theory as a guideline
for future experiments. Our goal was to identify those experimental settings
which are most likely to be sensitive to the four leading spin
polarizabilities of the proton and neutron. These quantities parameterize the
stiffness of the nucleon spin against electro-magnetically induced
deformations of definite multipolarity and non-zero frequency. Their energy
dependence gives profound insight into the dispersive behavior of the internal
degrees of freedom of the nucleon, caused by internal relaxation effects,
baryonic resonances and mesonic production thresholds, see also \cite{HGHP03,
   GH01} for details.

In the spin-averaged cross sections, Sect.~\ref{spinav}, we found significant
deviations between predictions with and without spin polarizabilities.
Therefore, spin-averaged experiments can contribute to a direct determination
of \textit{dynamical spin} polarizabilities from data, too.

In the polarized case, Sects.~\ref{sec:proton} and~\ref{sec:neutron}, we
considered configurations with a right-circularly polarized incident photon
and a polarized target nucleon, leaving the spins of the particles in the
final state undetected. We investigated the dependence of the cross sections
and asymmetries on the spin polarizabilities in two different spin
configurations: (i) nucleon spin parallel minus antiparallel to the photon
momentum, and (ii) perpendicular to it but still inside the reaction plane.
We noted a stronger sensitivity in the asymmetry $\Sigma_x$ of configuration
(ii) for both proton and neutron targets, than in case (i).  We found
furthermore that only two of the structure amplitudes, namely $A_1$ and $A_3$,
dominate all cross sections and asymmetries.

The spin polarizabilities give
usually the clearest signal for photon energies above 100 MeV, say around 
the pion production threshold ($\sim130$~MeV), where
most of the asymmetries also reach $\calO(1)$.  
In backward direction, the neutron asymmetries also show a strong
sensitivity on the physics of the $\Delta(1232)$ resonance, in addition to
contributions from the chiral pion cloud around the nucleon.

In general, the neutron asymmetries were found to be more sensitive to the
spin polarizabilities than the proton analogs. This is no surprise, since 
Compton scattering on
the charged proton is dominated by the pole amplitudes.  Besides the resulting
 minor sensitivity on the nucleon structure, another disadvantage of the proton
asymmetries is the small spin-averaged cross section
around the pion production threshold for small angles, which enhances
theoretical uncertainties. Again, we emphasize that up to leading-one-loop
order the only difference between proton and neutron is given by the pole
contributions, i.e.~the structure part of our amplitudes is the one of an
isoscalar nucleon.  Polarized cross sections, which are
calculated in this approximation, might therefore 
deviate from experimental results, especially for the neutron, where the pole
contributions are weak. Hints on 
important isovector contributions
are given in
\cite{Schu}, where the non-pion-pole contribution to the backward spin
polarizability $\gamma_\pi$ of the neutron was found about twice as large as
the corresponding proton value.

Contributions of the quadrupole polarizabilities turned out to be negligibly
small, as in the spin-averaged case \cite{HGHP03}. Therefore, like
spin-averaged observables, spin-polarized cross sections are well described by
only six energy-dependent functions: the two spin-independent and four spin
dipole polarizabilities. This lead us in Sect.~\ref{strucamp} to propose to
extract the energy dependence of the four spin polarizabilities of the
individual nucleons by a model-independent multipole expansion of the
structure amplitudes from a combination of polarized and unpolaried precision
experiments. Chiral Effective Field Theory with explicit $\Delta(1232)$ degrees
of freedom represents correctly the symmetries and low energy degrees of
freedom inside the nucleon in a model-independent way. One can therefore in a
first step accept our predictions of the energy dependence of the
polarizabilities as induced by dispersive effects and only fit their overall
normalization to experiment, thus obtaining their static values. At present,
only two linear combinations, $\gamma_0$ and $\gamma_\pi$, were measured
exerimentally on the proton at LEGS~\cite{Sandorfi} and
MAMI~\cite{Olmos01,Galler}, with partially conflicting values. 

Clearly, the lack of free neutron targets makes an extension of the work
presented here to light nuclei mandatory if experiments are to be interpreted,
especially in the light of feasibility studies on
Compton scattering on the deuteron and ${}^3$He at HI$\gamma$S/TUNL
\cite{Gao}. Work is therefore under way to consider spin polarized observables
on these configurations, systematically including all binding and pion
exchange current effects in an extension of Chiral Effective Field Theory to
light nuclei \cite{linear}.  Further investigations involving linearly
polarized photon beams are also in preparation \cite{linear}. We are therefore
confident that such future experiments will further our understanding of a
fundamental property of the nucleon: the response of its effective low energy
degrees of freedom, and in particular of its spin, in strong electric and
magnetic fields, 
as parameterized in the dipole polarizabilities.


\section*{Acknowledgments}

The authors acknowledge helpful discussions with H.~Gao and W.~Weise. We are
grateful to the ECT* in Trento for its hospitality. This work was supported in
part by the Bundesministerium f\"ur Forschung und Technologie, and by Deutsche
Forschungsgemeinschaft under contract GR1887/2-1 (HWG and RPH).

\newpage

\appendix
\section{Dominant Amplitudes at Low Energies}
\setcounter{equation}{0}
\label{App:A1A3}
\begin{figure}[!htb]
\begin{center}
  \includegraphics*[width=.82\textwidth]{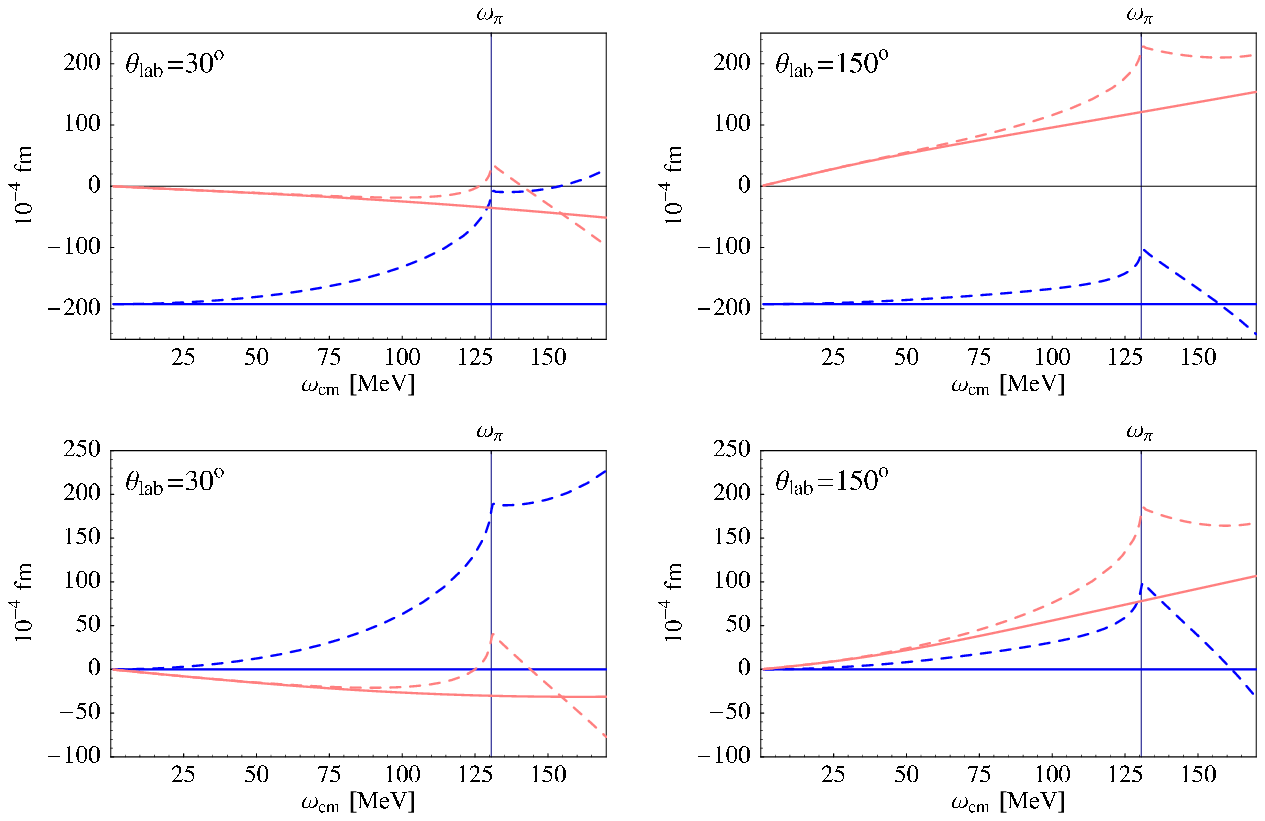}
\caption{$\mathcal{O}(\epsilon^3)$-SSE-results for the real parts of the 
  amplitudes $A_1$ (blue/dark) and $A_3$ (red/light); the dashed line is the
  full $\mathcal{O}(\epsilon^3)$-SSE-result, the solid line the third order
  pole contribution, given in Eqs.~(\ref{eq:pole}, \ref{eq:Bornn}); the upper
  (lower) panels show the proton (neutron) results. Recall that for the
  neutron, $A_1^\mathrm{pole}=0$.}
\label{A1A3}
\end{center}
\end{figure}

In Fig.~\ref{A1A3}, we plot $A_1$ and $A_3$ at forward and backward angles, as
we found in Sects.~\ref{spinav},~\ref{sec:proton} and~\ref{sec:neutron} many 
low energy features in
our results for polarized and unpolarized cross sections that can be explained
by considering only those two amplitudes.  The upper two panels in
Fig.~\ref{A1A3} show the plots for the proton in extreme forward and backward
direction, the lower two the ones for the neutron.  $A_1$ is sketched
dark/blue, $A_3$ light/red. The solid lines are the pole contributions from
Eq.~(\ref{eq:pole}) and Eq.~(\ref{eq:Bornn}), the dashed lines are the full
(i.e.~up to $l=2$) $\mathcal{O}(\epsilon^3)$-results.  The amplitudes of the
uncharged neutron vanish for $\omega=0$ while the static value of $A_1^p$ is
given by the Thomson limit.

The clear cusp structure arises from $\alpha_{E1}(\omega)$ in $A_1$ and from
$\gamma_{E1E1}(\omega)$ in $A_3$, cf. Eq.~(\ref{eq:strucamp}) and
\cite{HGHP03}.


\newpage



\end{document}